\documentclass[pre,preprint]{revtex4}
\usepackage{amsmath,amssymb}
\usepackage{epsfig}
\usepackage{graphics}
\usepackage{color}
\usepackage{psfrag}

\begin{document}

\newlength{\mylen}
\setlength{\mylen}{\textwidth}
\addtolength{\mylen}{-1cm}
\newcommand{\bea}{\begin{eqnarray}}
\newcommand{\eea}{\end{eqnarray}}
\newcommand{\spaceint}[2]{\int_{#1} d^3 #2 \;}
\newcommand{\vect}[1]{\mathbf{#1}}
\newcommand{\vat}{V^{\rm att}}
\newcommand{\di}{\displaystyle}
\newcommand{\req}{\rho_{\rm eq}}
\newcommand{\mex}{\mu^{\rm ex}}
\newcommand{\rp}{r_{||}}
\newcommand{\vrp}{{\vect r}_{||}}
\newcommand{\qp}{q_{||}}
\newcommand{\vqp}{{\vect q}_{||}}
\newcommand{\ft}{{\rm FT}\;}
\newcommand{\bt}[1]{{\rm HT}_{#1}\;}
\newcommand{\tx}{\textcolor{blue}}

\title{Hard sphere fluids in annular wedges: density distributions and depletion potentials }

\author{V. Bo{\c{t}}an}
\author{F. Pesth}
\author{T. Schilling}
\author{M. Oettel}
\email{oettelm@uni-mainz.de}
\affiliation{Johannes--Gutenberg--Universit\"at Mainz, Institut f{\"ur} Physik,
  WA 331, D--55099 Mainz, Germany,}

\begin{abstract}
 We analyze the density distribution and the adsorption of solvent hard spheres in an annular slit 
 formed by two large solute 
 spheres or a large solute and a wall at close distances 
 by means of fundamental measure density functional theory, anisotropic
 integral equations and simulations. We find that the main features of the density 
 distribution in the slit are described by an effective, two--dimensional system of disks
 in the vicinity of a central obstacle.
 This has an immediate consequence for the depletion force between the solutes (or the wall and 
 the solute), since the latter receives a strong line--tension contribution due to
 the adsorption of the effective disks at the circumference of the central obstacle.
 For large solute--solvent size ratios, the resulting depletion force has a straightforward 
 geometrical interpretation which gives a precise ``colloidal" limit for the depletion 
 interaction. For intermediate size ratios 5 \dots 10 and high solvent packing fractions larger than 0.4,
 the explicit density functional results show a deep attractive well for the depletion potential
 at solute contact, possibly indicating demixing in a binary mixture at low solute
 and high solvent packing fraction.   
\end{abstract}

\maketitle

\section{Introduction}

\label{sec:intro}

The equilibrium statistical theory of inhomogeneous fluids whose foundations were laid out in the
1960's \cite{Mor60,Ste64} is a well--studied subject which has been developed since by a fruitful interplay between 
simulations and theory. For the simplest molecular model, the hard--body fluid mixture, this has led to
the development of powerful, geometry--based density functionals \cite{Ros89,Rot02,Yu02,Han06,Mal06,Han09} 
which accurately describe
adsorption phenomena, phase transitions (such as freezing for pure hard spheres or
demixing for entropic colloid--polymer mixtures) and molecular layering near obstacles.
Such a class of density functionals has not been found yet for fluids with attractions, however,
significant progess has been achieved in the description of bulk correlation functions
and phase diagrams through the method of integral equations \cite{Pin02,Par08,Aya09}.

Strong inhomogeneities occur if fluids are confined
on molecular scales, a topic which enjoys continuous interest \cite{Tru08,Jun08}. For example, 
the packing of molecules between 
parallel walls leads to oscillating forces between them which can be measured experimentally
and determined theoretically \cite{Isr}. With the development of preparational techniques for colloids and
their mixtures, the ``molecular" length scale has been lifted to the range of several nm up
to $\mu$m which even allows the direct observation of modulated density profiles through microscopy
besides the measurement of resulting forces on the walls. Furthermore, in the colloidal domain the
possibility to tailor the interparticle interactions to a certain degree allows the
close realization of some of the favourite models for simple fluids fancied by theorists,
such as e.g. hard spheres \cite{Dul06}.    
This has opened the route to quantitative comparisons between experiment, theory and simulations.

In an asymmetric colloidal mixture with small ``solvent" particles and at least one species
of larger solute particles, the phenomenon of solvent--mediated, effective interactions between
the solute particles may give rise to various phase separation phenomena \cite{Lik01}. 
From a theoretical point of view, these effective interactions are
interesting since they facilitate the description of mixtures in terms of an
equivalent theory for one species interacting by an effective potential
\cite{McMillan45,Dij99}.
If the solvent particles possess hard (or at least steeply repulsive) cores, they are excluded
from the region between two solute particles if the latter are separated by less than one solvent
diameter. This gives rise to strong depletion forces whose understanding is crucial
for the concept of an effective theory containing only solute degrees of freedom. 
Since the magnitude of the depletion force is directly linked to the solvent density distribution
around the solutes, the quantitative investigation of the solvent confined between solute
particles appears to be important. The confinement becomes rather extreme for large asymmetry
between solute and solvent (see below).
 
\begin{figure}
 \begin{center}
  \epsfig{file=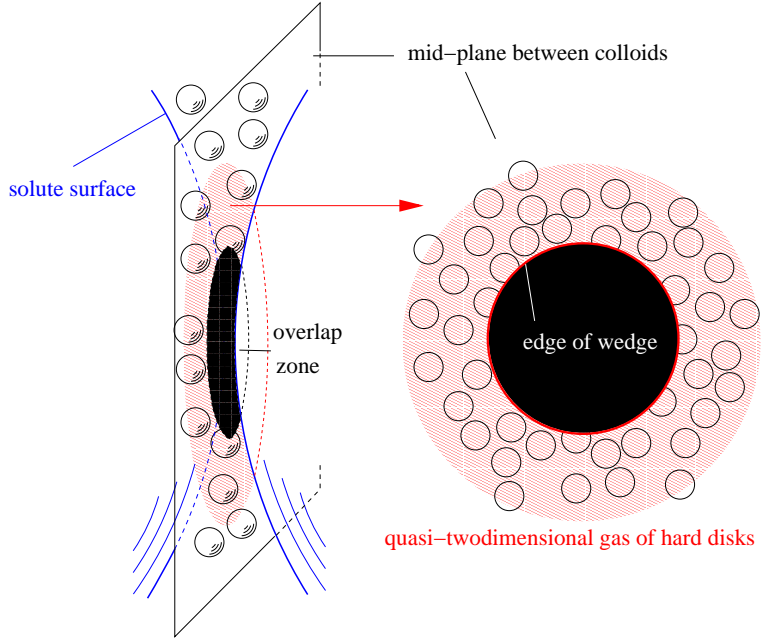, width=10cm}
 \end{center}
 \caption{(color online) The annular wedge which is formed between two large solute particles for separations
  $h\le \sigma$. Black areas denote domains which are forbidden for the centers of 
  the solvent spheres.}
 \label{fig:slit}
\end{figure}

In the following we want to concentrate on the idealized system of additive hard
spheres and in particular on the effective interaction between two solute particles in 
solvent, i.e. the case of infinite dilution of solute particles in the colloidal mixture.
The case of the interaction of one  solute particle with a wall is a special case in that
the radius of the other solute particle goes to infinity.  
For the case of hard spheres (solvent  
diameter $\sigma=2R_s$, solute radius $R_b$ such that the size ratio is $\alpha=R_b/R_s$ and
$R=R_b+R_s$ is the radius of the exclusion sphere for solvent centres around a solute sphere) 
it has been found that the theoretical description using ``bulk" methods becomes
increasingly inaccurate for size ratios $\alpha \agt 5$ and solvent densities 
$\rho_s^* = \rho_s \sigma^3 \agt 0.6$, when compared to simulations.
 Here, the terminus ``bulk" methods refers to
methods which determine the bulk pair correlation function between the colloids, $g_{bb}(r)$,
from which the depletion potential is obtained as $\beta W = - \ln g_{bb}$. 
Bulk integral equation (IE) methods \cite{Att90} and the ``insertion" trick in density functional
theory (DFT) \cite{Rot00,Amo01,Aya05,Amo07} fall into this category. For larger size ratio, one would expect 
that the well--known Derjaguin approximation \cite{Der34,Hen02,Oet04} 
becomes accurate very quickly. Interestingly, however,
both the mentioned bulk methods and simulations deviate significantly from the Derjaguin 
approximation (besides disagreeing with each other) for size ratios 10 and solvent
densities $\rho_s^*=0.6 \dots 0.7$ \cite{Bib96,Dic97,Oet04} when the surface--to--surface 
separation $h$ between
the solute particles is close to $\sigma$ ($h \alt \sigma$, i.e. at the onset of the depletion 
region where the solvent particles are ``squeezed out" between the solute particles).   
One notices that for $h\le\sigma$ the solute particles form an annular wedge 
with a sharp edge which restricts the 
solvent particles to quasi--2d motion (see Fig.~\ref{fig:slit}). 
Taking this observation into account, the 
phenomenological analysis of Ref.~\cite{Oet04} 
predicted that the solvent adsorption at this edge leads to a line contribution to
the depletion potential which is proportional to the circumference of the circular egde and thus
to $R^{1/2}$. Such a term in the depletion potential causes a very slow approach to the
Derjaguin limit for large solutes (the Derjaguin approximation describes the depletion potential
essentially by volume and area terms of the overlap of the exclusion spheres pertaining
to the solutes \cite{Hen02,Oet04}, see below). In this way, a generalized Derjaguin approximation 
serving as a new ``colloidal'' limit can be formulated
which turns out to have a far more general meaning \cite{Oet09} than anticipated. 
Using the concept of morphological (morphometric) (morphometric) (morphometric) (morphometric) (morphometric) (morphometric) (morphometric) (morphometric) thermodynamics introduced in Ref.~\cite{Koe04}, one finds that
the insertion free energy of two solute particles (and thus the depletion potential between
them) only depends on the volume, surface area, and the integrated mean and Gaussian
curvatures of the solvent accessible surface around the two solutes. The coefficients of these
four terms depend on the solvent density {\em but not on the specific type of surface}.
In this manner, the phenomenological
line tension of Ref.~\cite{Oet04} is related to the general coefficient of mean curvature
of the hard--sphere fluid \cite{Oet09}, and the Derjaguin approximation is equivalent to the morphometric
analysis restricted to volume and surface area terms.  

Since the depletion force in the hard--sphere system is directly linked to an integral
over the solvent contact density on one solute (see Eqs.~(\ref{eq:fdef}) and (\ref{eq:fwdef}) below), one should be
able to connect the morphometric approach to features of the solvent density profile
in the annular wedge. This is a strong motivation for us to explicitly determine the wedge density 
profile. We will do so by means of density functional theory and anisotropic integral equations,
and compare it to results of Monte--Carlo (MC) simulations for selected parameters.      
For the intermediate densities $\rho_s^*=0.6$ and $0.7$ and various size ratios between solute and
solvent results for the depletion force from such explicit DFT calculations were
already presented in Ref.~\cite{Oet09}, and good agreement with the morphometric depletion force
was obtained for size ratios between 5 and 40. Here, we will present a detailed analysis of the
wedge density profiles and consider also higher densities. We will give a thorough comparison
to recent results from MC simulations which have been obtained for the wall--solute interaction
at a solvent density $\rho_s^*=0.764$ ($\eta_s=(\pi/6)\rho_s \sigma^3=0.4$) 
and size ratios between 10 and 100 \cite{Her06,Her07,Her08}. 

The paper is structured as follows. In Sec.~\ref{sec:theory} we introduce the basic notions of density
functional and integral equation theory and present a short description of the Monte Carlo simulations
employed here. In Sec.~\ref{sec:wall_solute} we analyze in detail the density profiles and the 
corresponding depletion forces for the wall--sphere geometry for the particular solvent density
$\rho_s^*=0.764$. This geometry permits explicit 
calculations up to solute--solvent size ratios 100 and a test of the ``colloidal limit" of the 
morphometric depletion force. Sec.~\ref{sec:solute_solute} gives an analysis of the depletion  
force and potential in the sphere--sphere geometry (corresponding to the important case of the 
effective interaction in a dilute mixture) for intermediate size ratios 5 and 10 and higher solvent
densities $\rho_s^*=$ 0.8 \dots 0.9. 

\section{Theory and methods}

\label{sec:theory}

\begin{figure}
   (a)\epsfig{file=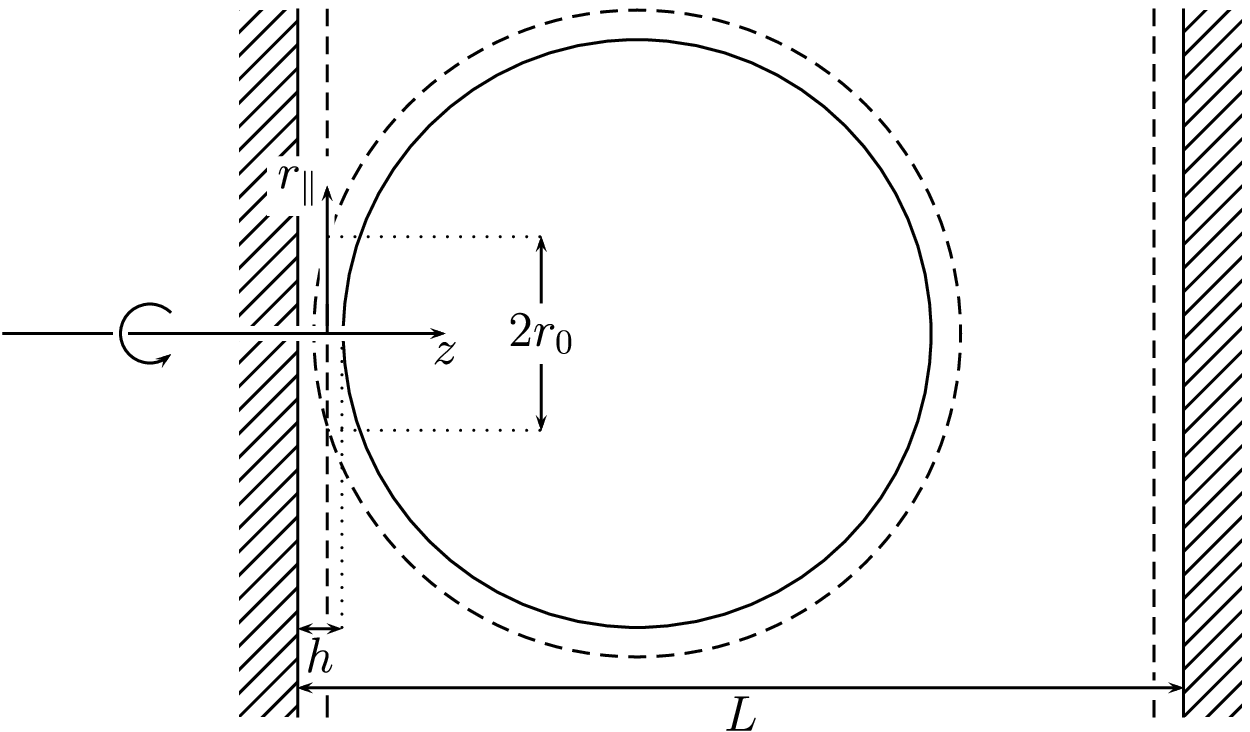,width=8cm} \\[10mm]
   (b)\epsfig{file=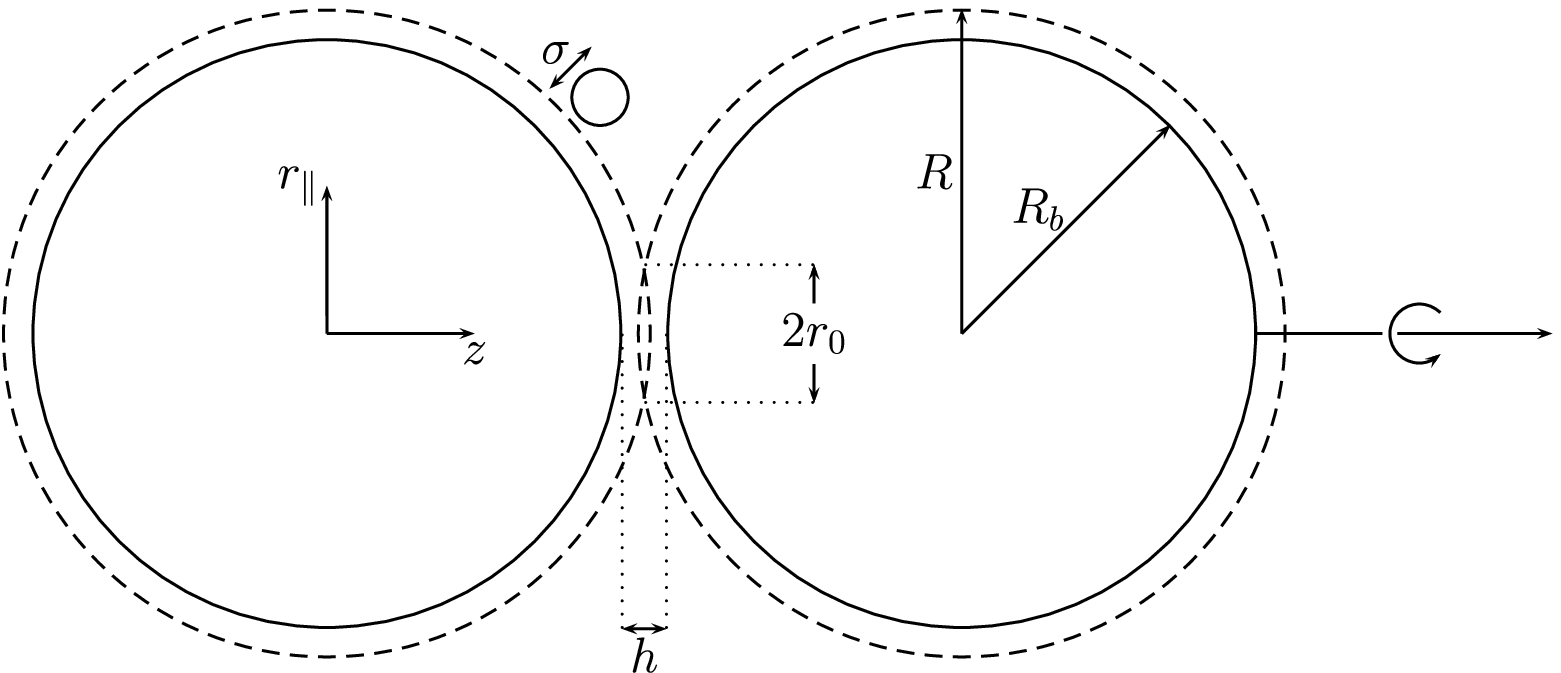,width=10cm} \\
 \caption{View of the geometric configurations used in this work. (a) Solute sphere of radius $R_b$ immersed in a 
solvent--filled slit of width $L$. Only the density profile in the annular wedge between the left wall
and the solute is of interest since it determines the depletion force between solute and one wall.
Note that one can also determine the slit density profile and the corresponding depletion force
for $2R_b > L$ (i.e. when the solute does not fit into the slit) as long as $L$ is large enough
that the correlations from the right wall do not reach into the annular wedge. 
(b) Two solute spheres of radius $R_b$ at distance $h$ immersed in bulk solvent.    
For both setups, the solvent sphere diameter is given by $\sigma$, and the radius of the exclusion sphere
around a solute particle is given by $R=R_b+\sigma/2$.}
 \label{fig:setup}
\end{figure}

We consider two scenarios (see Fig.~\ref{fig:setup}): (a) One hard solute sphere
immersed in the hard sphere solvent of density $\rho_s$ confined to a slit, created by two hard walls at distance $L$. The 
surface--to--surface distance between one wall and the solute sphere is denoted by $h$,
furthermore $L \gg h$ such that the correlations from one wall do not influence the solute 
interaction with the other wall. 
(b) Two hard solute spheres immersed in the bulk solvent
spheres with surface--to--surface distance $h$. The solvent accessible surface is given by the dashed lines
in Fig.~\ref{fig:setup}, thereby one sees that for $h \le \sigma$ the two solutes (or the solute and one wall)
form an annular wedge in which the solvent adsorbs. The solvent density profile 
$\rho(\vect r) \equiv \rho(\rp,z)$ depends only on $z$, the coordinate on the symmetry axis
and the distance $\rp$ to the symmetry axis. In the latter case (b), when the first colloid is centered
at the origin and the second one at $z=2R_b+h$,  the total force $f(h)$ on the first colloid is 
obtained by integrating the force $-\nabla u_{bs}(\vect r)$ ($u_{bs}$ is the solute--solvent potential) 
between the solute and one solvent  
sphere over the density distribution $\rho(\vect r)$:
\bea
   \label{eq:fdef}
  \beta f(h) &=& - \int d^3 r \;\rho(\vect r)\; \vect e_z \cdot \nabla u_{bs}(\vect r) \\
  \nonumber
  & =& 2\pi R^2\;\int_{-1}^1 d(\cos\theta) \cos\theta\;
    \rho(\vect r) \;, \\
   & & \quad \left[ |\vect r|=R\;, \quad \cos\theta = \hat {\vect r} \cdot \vect e_z
    \right] \;. \nonumber
\eea   
Thus the total depletion force reduces to an integral over essentially the contact density on 
the exclusion sphere around the colloid. This follows from 
$\exp(-\beta u_{bs})\beta \nabla u_{bs} = - \hat {\vect r} \delta ( |\vect r|-R) $ and the observation
that $\rho(\vect r)\exp(\beta u_{bs})$ is continuous across the exclusion sphere surface.
In case (a), the force on the colloid is the negative of the excess force $f_w$ on the wall
and the excess force is determined by the total force on the wall minus the contribution from
the bulk solvent pressure $p$. Through the wall theorem, the latter is given by $\beta p =\rho_w$
where $\rho_w$ is the contact density of the solvent at a single wall.
By an argument similar to the above one and putting the exclusion surface of the
wall at $z=0$, $f_w$ is determined as
\bea
  \label{eq:fwdef}
  \beta f_w(h) &=& -2\pi \int_0^\infty \rp\,d\rp\;(\rho(\rp,z=0) - \rho_w) \;.
\eea
Thus, $f_w$ is equivalent to the excess adsorption at the wall which makes it somewhat easier to
determine in MC simulations than $f$ \cite{Her07}. 
 
\subsection{Density functional theory}

The equilibrium solvent density profile $\rho(\vect r) \equiv \req(\vect r)$
can be  determined directly from the basic equations 
of density functional theory. The grand potential functional is given by 
\bea
 \label{eq:omegadef}
  \Omega[\rho] &=& {\cal F}^{\rm id}[\rho] +  {\cal F}^{\rm ex}[\rho] - \int
   d\vect r(\mu - V^{\rm ext}(\vect r)) \;,
\eea 
where ${\cal F}^{\rm id}$ and ${\cal F}^{\rm ex}$ denote the ideal and excess free
energy functionals of the solvent. The solvent chemical potential is denoted by $\mu$ and
the solute(s) and/or the walls define the external potential $V^{\rm ext}$. 
The ideal part of the free energy is given by
\bea
 \label{eq:fid}
 \beta {\cal F}^{\rm id} &=& \int d\vect r \,\rho(\vect r) 
                          \left(\ln(\rho(\vect r) \Lambda^3)-1 \right)\;, 
\eea
with $\Lambda$ denoting the de--Broglie wavelength.
The equilibrium density profile $\req(\vect r)$ for the solvent at
chemical potential $\mu=\beta^{-1}\ln(\rho_s\, \Lambda^3) + \mex$
(corresponding to the bulk density $\rho_s$)
is found by minimizing the grand potential in Eq.~(\ref{eq:omegadef}):
\bea
 \label{eq:rhoeq_main}
  \ln \frac{\req(\vect r)}{\rho_s} + \beta V^{\rm ext}(\vect r) =
   -\beta\frac{\delta {\cal F}^{\rm ex}[\req]} {\delta \rho(\vect r)}+ \beta\mex  \;. 
\eea
For an explicit solution, it is necessary to specify the excess part of the
free energy.
Here we employ two functionals of fundamental
measure type (FMT). These are the original Rosenfeld functional (FMT--RF) \cite{Ros89} and the
White Bear functional (FMT--WB, with Mark I from Refs.~\cite{Rot02,Yu02} and Mark II from
Ref.~\cite{Han06}), for another closely related variant see Ref.~\cite{Mal06}. 
It has been demonstrated that FMT gives very precise
density profiles also for high densities of the hard sphere fluid in various circumstances. 
Furthermore, the White Bear II functional possesses a high degree of self--consistency
with regard to scaled--particle considerations \cite{Han06}.
However, we do not consider the tensor--weight modifications of these 
functionals which are necessary to obtain a correct description of the liquid--solid
transition \cite{Tar00} and are of higher consistency in confining situations which reduce
the dimensionality of the system (``dimensional crossover", see Ref.~\cite{Tar97,Ros97}). 
This might be an issue in some circumstances (see below).

Th explicit forms of the excess free energy are given in
App.~\ref{app:fmt_min}. 
Numerically, due to the non--local nature of ${\cal F}^{\rm ex}$, 
Eq.~(\ref{eq:rhoeq_main}) corresponds to an integral equation   
for the density profile depending on the two variables $\rp$ and $z$.
The sharpness of the annular wedge in the solute--solute and the solute--wall
problems for high size ratios necessitates rather fine gridding (which makes
the calculation of ${\cal F}^{\rm ex}$ 
and $\delta {\cal F}^{\rm ex}/\delta \rho$ very time--consuming) and introduces
an unusual slowing--down of standard iteration procedures of Picard type.
To overcome these difficulties, we made use of fast Hankel transform techniques and
more efficient iteration procedures. These are described in detail in     
App.~\ref{app:fmt_min} as well.  

We remark that there have been earlier attempts to obtain explicit density distributions
using DFT around two fixed solutes and thereby to extract depletion forces 
\cite{Arc05,Ego04,Gou01}. 
In Ref.~\cite{Ego04} this was done using the simple Tarazona I functional for hard spheres \cite{Tar84}, 
and considerations
were limited to a small solvent packing fraction of 0.1 and solute--solvent size ratios of 5. In
Ref.~\cite{Gou01}, a minimization of FMT--RF was carried out using a real--space technique
for solvent densities up to $\rho_s^*=0.6$ and size ratios 5. In that respect, the present technique
is superior in that it allows us to present solutions for  solvent densities up to $\rho_s^*=0.9$ and 
size ratios up to 100. 

\subsection{Integral equations}

The central objects within the theory of integral equations are the correlation
functions on the one and two--particle level in the solvent.
We are interested in the two--particle correlation functions in the presence of
an external background potential (one fixed object, solute or wall), from which the explicit solvent density profiles
around the two fixed objects (solute--solute or solute--wall) follow (see below). 
This is usually referred to as the method of inhomogeneous or anisotropic integral equations, 
first employed for Lennard--Jones fluids on solid substrates
\cite{Hillebrand84,Bruno87} and for hard--spheres (HS) fluids in
contact with a single hard wall in Ref.~\cite{Plischke86}. Liquids
confined between two parallel walls (a planar slit) have been
extensively studied by Kjellander and Sarman \cite{Kjellander88,Kjellander90,Kjellander91}. 

For an arbitrary background potential $V(\vect r)$ which corresponds
to an equilibrium background density profile for the solvent $\rho_V(\vect r)$, the pair correlation
function $g_{ij}(\vect r, \vect r_0) = h_{ij} (\vect r, \vect r_0) + 1$ describes the
normalized probability to find a particle of species $i$ at position $\vect r$ if another 
particle of species $j$ is fixed at position $\vect r_0$. For our system, the species
index is either $s$ (solvent particle) or $b$ (big solute particle).
 In the solute--solute
case, $V$ is given by the potential of one solute particle, whereas in the
solute--wall case it is the potential of the wall. Then the equilibrium density profile
discussed in the previous subsection is related to the pair correlation function through
$\rho(\vect r) = \rho_V(\vect r)\, g_{sb}(\vect r, \vect r_0)$ ($\vect r_0$ specifies the position of the
(other) solute particle). The depletion force follows then through Eqs.~(\ref{eq:fdef}) and (\ref{eq:fwdef}).    

The corresponding direct correlation functions of second order $c_{ij}(\vect r, \vect r_0)$ are related to
$h_{ij}$ through the inhomogeneous Ornstein--Zernike (OZ) equations
\bea
  h_{ij}(\vect r, \vect r_0) - c_{ij}(\vect r, \vect r_0) = \sum_{k=b,s}
  \int d\vect r' \rho_{V,k}(\vect r') h_{ik}(\vect r, \vect r') c_{kj}(\vect r', \vect r_0) 
\eea 
In the dilute limit for the solute particles which we consider here, the background density 
$\rho_{V,b}$ for the solutes is zero, therefore the OZ equations reduce to
\bea
 \label{eq:ozss}
  h_{ss}(\vect r, \vect r_0) - c_{ss}(\vect r, \vect r_0) & = &
 \int d\vect r' \rho_{V}(\vect r') h_{ss}(\vect r, \vect r') c_{ss}(\vect r', \vect r_0) \;, \\
 \label{eq:ozbs}
   h_{bs}(\vect r, \vect r_0) - c_{bs}(\vect r, \vect r_0) & = &
    \int d\vect r' \rho_{V}(\vect r')  h_{bs}(\vect r, \vect r') c_{ss}(\vect r', \vect r_0) \;.
\eea
The background density profile is linked to $c_{ss}$ through the Lovett--Mou--Buff--Wertheim equation \cite{LMBW}:
\bea
 \label{eq:lmbw}
  \nabla \rho_V(\vect r) = -\beta \rho_V(\vect r) \nabla V(\vect r) +
  \rho_V(r) \int d\vect r' c_{ss}(\vect r,\vect r') \nabla\rho_V(\vect r')\;.
\eea

A third set of equations is necessary to close the system of equations.
The diagrammatic analysis of Ref.~\cite{Mor60} provides the general form of this closure which
reads:
\bea
 \label{eq:cl}
  \ln g_{ij} (\vect r, \vect r_0) + \beta u_{ij}(\vect r-\vect r_0) &=& 
  h_{ij}(\vect r, \vect r_0) - c_{ij}(\vect r, \vect r_0) - b_{ij}(\vect r, \vect r_0) \;,
\eea 
where $u_{ij}$ are the pair potentials in the solute--solvent mixture and $b_{ij}$ denote the
bridge functions specified by a certain class of diagrams which, however, can not be resummed in
a closed form.
For practical applications, these bridge 
functions need to be specified  in terms of $h_{ij}$ and $c_{ij}$ to arrive at a closed system
of equations. Among the variety of empirical forms devised for this connection we mention those
which, in our opinion, rely on somewhat more general arguments. 
These are the venerable hypernetted chain (HNC), Percus--Yevick (PY) closure and the mean 
spherical approximation (MSA) which can be derived
by systematic diagrammatic arguments \cite{HMD06}, and  the reference HNC closure which
introduces  a  suitable reference system for obtaining
$b_{ij}$ with subsequent free energy minimization \cite{Lad83}. 
(For bulk properties of liquids, the self--consistent Ornstein--Zernike approximation (SCOZA) \cite{Pin02} and
the hierarchical reference theory (HRT) \cite{Par08} are very successful. They rely on a closure of MSA 
type but it appears to be difficult in generalizing them to inhomogeneous
situations such as considered here.) In calculations, we considered the closures 
\bea
 \label{eq:py}
  b_{ij} & = & \gamma_{ij} - \ln(1+\gamma_{ij})  \qquad \mbox{(PY)}\; ,  \\
 \label{eq:ry}
b_{ij} &= &\gamma_{ij}-\ln\left [
1+\frac{\exp\{[1-\exp(-\xi_{ij}r)]\gamma_{ij} \}-1}{1-\exp(-\xi_{ij}r)}\right] \qquad \mbox{(RY)}\; , \\
\label{eq:vm}
b_{ij} &= &\frac{1}{2}\frac{\gamma_{ij}^2 }{1+\alpha_{ij}\gamma_{ij}} \qquad \mbox{(MV)}\; . 
\eea
where $\gamma_{ij}=h_{ij}-c_{ij}$. 
The Percus--Yevick (PY) approximation Eq.~(\ref{eq:py})
is exactly solvable in the bulk case even for the
muticomponent HS fluid \cite{Lebowitz}. However, beyond providing
good qualitative behavior PY does not produce precise quantitative
results in general, since it fails at contact, where
the value of pair distribution function is too small. 
For bulk systems, PY generates a noticeable thermodynamic inconsistency, i.e. a discrepancy between
different routes to the equation of state. 
To overcome these deficiencies a
refined approximation to the closure, which interpolates between HNC ($b_{ij}=0$)
and PY closures was suggested by Rogers and Young (RY) \cite{RY}, see Eq.~(\ref{eq:ry}) where 
the $\xi_{ij}$ are adjustable parameters, which can be  found from the thermodynamic
consistency requirement \cite{RY}.
An even more suitable for the asymmetric HS mixtures
variant of the modified Verlet (MV) closure was suggested in
Ref.~\cite{Henderson96}, see Eq.~(\ref{eq:vm}). Here the parameters $\alpha_{ij}$ are chosen to satisfy the exact
relation between $b_{ij}(0)$ and the third virial coefficient at low
densities.  For the choice of the auxiliary parameters $\xi_{ij}$ and $\alpha_{ij}$ in the present work, see App.~\ref{app:ie}

We have solved the set of equations (\ref{eq:ozss})--(\ref{eq:vm}) for the solute--wall case where we could
employ similar numerical methods as in the numerical treatment of density functional theory. (This is not
possible for the solute-solute case, for an efficient method applicable in this case, see
Ref.~\cite{Att89}.) In the solute--wall case, the background density profile
$\rho_V(\vect r) \equiv \rho_V(z)$ depends only on the distance $z$ from the wall. The two--particle
correlation functions depend on the $z$--coordinates of the two particles individually and the
difference in the radial coordinates: $g_{ij}(\vect r, \vect r_0) \equiv g_{ij}(z,z_0,\rp-r_{||,0})$. Thus, the Ornstein--Zernike equations (\ref{eq:ozss}) and (\ref{eq:ozbs}) become
matrix equations for the correlation functions in the $z$--coordinates, and are diagonal 
for the Hankel transforms of the correlation functions in the $\rp$--coordinates. For more
details on the numerical procedure, see App.~\ref{app:ie}.

\subsection{Simulation Details}

\begin{figure}[ht]
 \begin{center}
  \epsfig{file=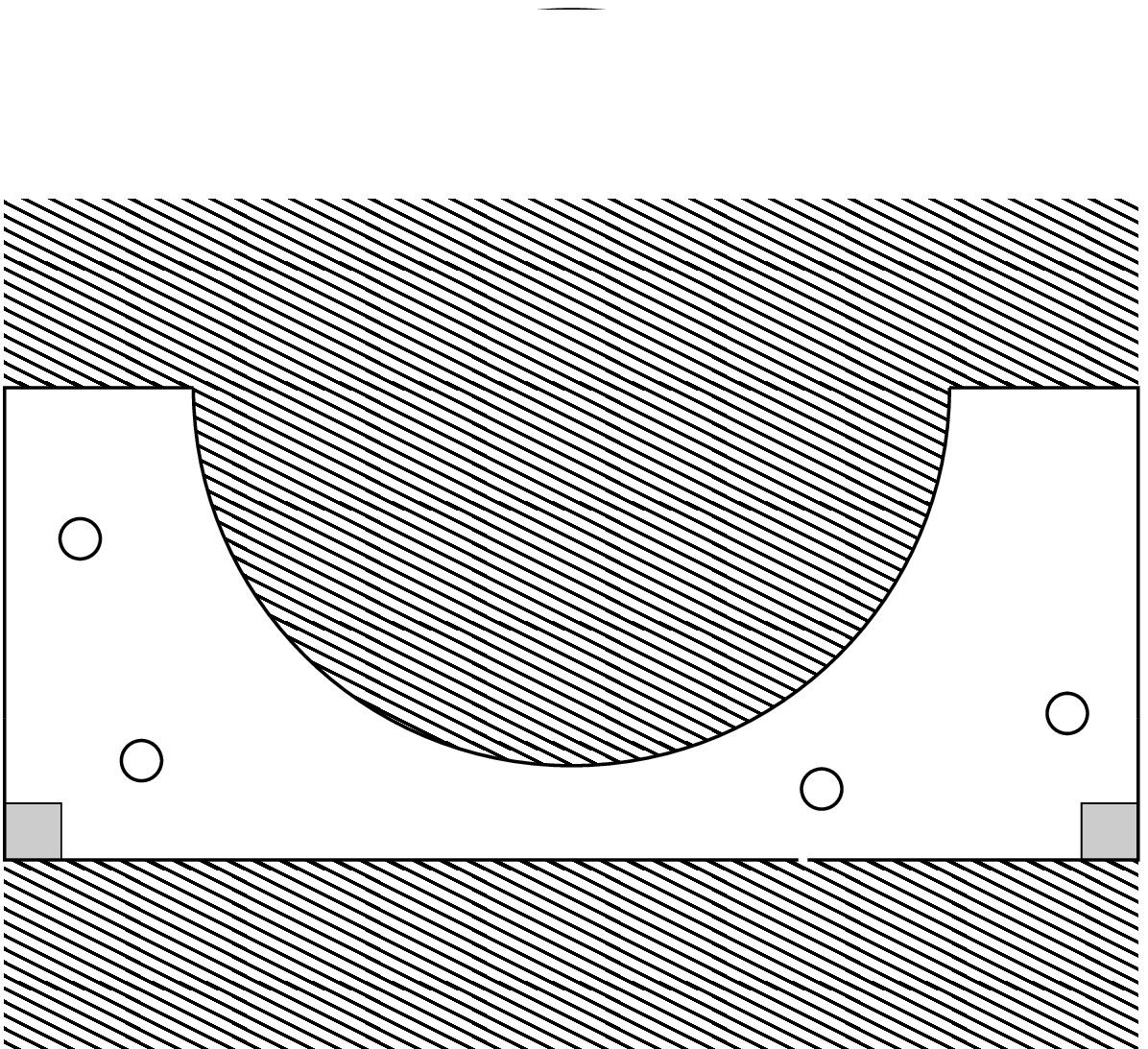,width=7cm}
  \end{center}
\caption{Sketch of the simulation box. Hard walls are
  marked by hatches. The remaining boundaries were treated as periodic. The
  grey squares mark the regions in which the density at wall contact
was sampled.} 
\label{SimuBox}
\end{figure}

The Monte Carlo simulations for the wall--solute system were performed at fixed particle number and 
volume of the simulation box. The upper boundary of the box was given by a 
large hard sphere, the lower boundary by a planar hard wall 
(see hatched regions in Fig.~\ref{SimuBox}). All remaining boundaries 
were treated as 
periodic. We sampled the configuration space of the small spheres by
standard single particle translational moves. In order to impose the asymptotic solvent 
bulk density $\rho_s^*=0.764$ ($\eta_s=0.4$), the concentration
of small spheres in the box was set such that the density at contact with the
hard wall far away from the wedge (i.~e.~in the region marked by grey 
squares in Fig.~\ref{SimuBox}, averaged over a depth 0.02 $\sigma$) 
settled to $\bar\rho_{\rm Wall} = 4.88$ 
within 1\% error. 
This value $\bar\rho_{\rm Wall} = 4.88$
was obtained from the density profile of hard spheres at a hard wall at the bulk
packing fraction 0.4, calculated with
FMT--WBII which is very accurate.
System sizes ranged from 1800
particles for $R_b=10\;\sigma$ to 8000 particles for $R_b=25\;\sigma$.  
In order to
access the configurations inside the narrow part of the wedge with sufficient
accuracy, large numbers of Monte Carlo sweeps were required.     
 We equilibrated the systems for $5 \times 10^5$ MC sweeps 
(i.~e.~attempted moves per particle) each. For data acquisition, we performed 
between $10^7$ and $10^8$ sweeps (depending on the parameters $h$ and $R_b$) and
averaged over $10^6$ to $10^7$ samples. Note that from the simulations we obtained
only the wedge density profiles (see Figs.~\ref{fig:rho_wall_20}--\ref{fig:h1scaling} below) 
but did not attempt to obtain the
depletion force from Eq.~(\ref{eq:fwdef}) where one needs an excess adsorption integral  
at the wall. This would require considerably more sweeps \cite{Her07,Her08}, see also
Fig.~\ref{fig:hforce} below for the statistical errors of the simulated depletion force
according to Ref.~\cite{Her07}.

\section{Results}

\subsection{Wall--solute interaction}

\label{sec:wall_solute}

In this section, we analyze case (a), the problem of a solute sphere close to a wall. 
Previous simulation work \cite{Her06,Her07,Her08} has concentrated on the particular 
state point $\rho_s^*=0.764$ (packing fraction $\eta_s=0.4$), therefore we present explicit
results for the density profiles also for this state point to facilitate comparison.   

\subsubsection{Density profiles}
 
\begin{figure}
 \begin{center}
   \epsfig{file=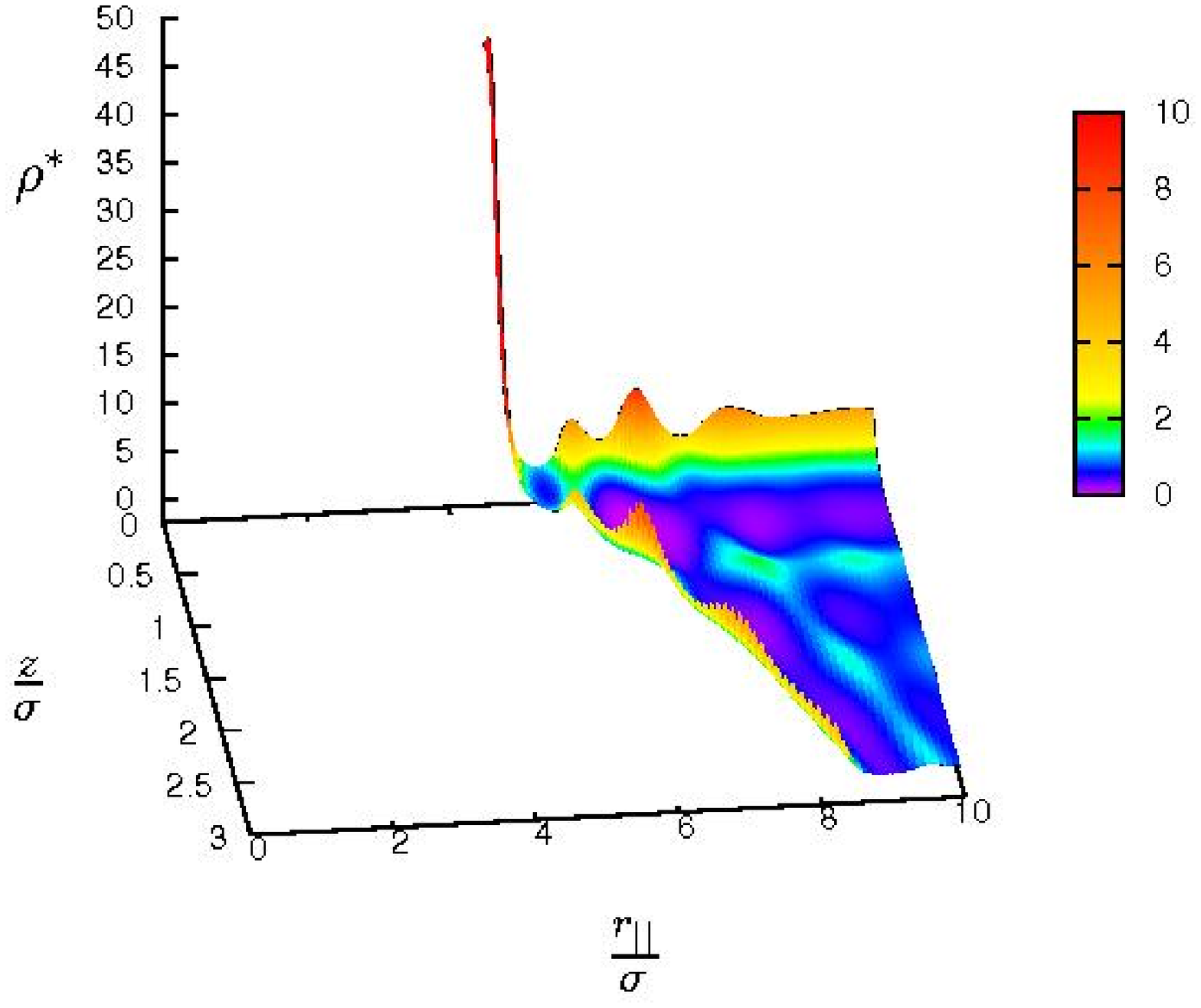,width=9cm} 
   \epsfig{file=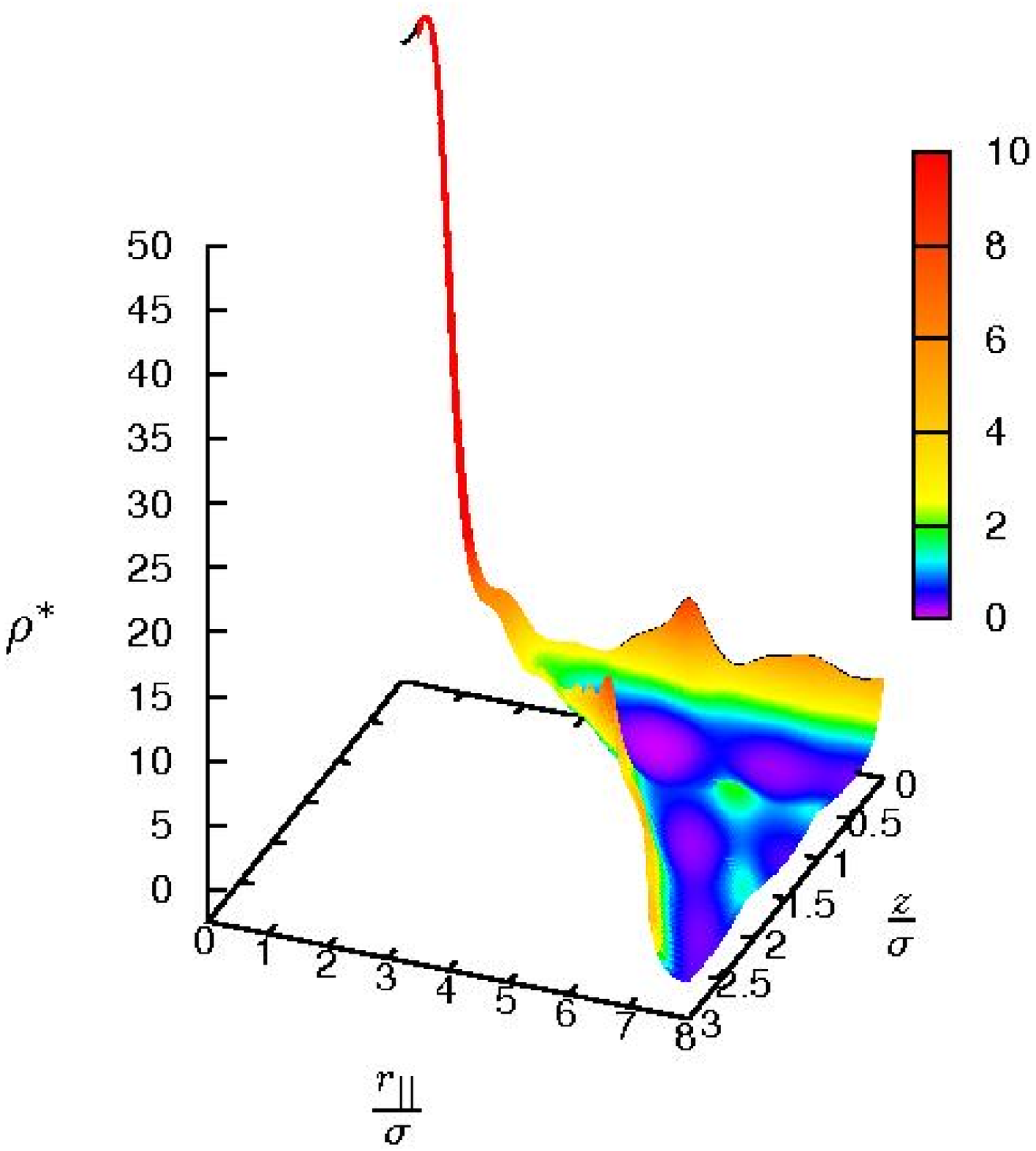, width=7cm} 
 \end{center}
 \caption{(color online) Density profiles from FMT--WBII of solvent spheres confined to the annular wedge between a solute
sphere (size ratio $R_b/R_s=20$) and a wall. The bulk solvent density is $\rho_s^*=0.764$, and the wall separation
$L=26\,\sigma$ (see Fig.~\ref{fig:setup} (a)). Grid resolution was equidistant  in $z$ with
 $\Delta z= 0.002$ $\sigma$ and
equidistant in $x=\ln (\rp/\sigma)$ with $\Delta x = 0.005$ (see App.~\ref{sec:fht}).  
Left panel: solute--wall distance $h=0$, right panel:
$h=\sigma$.  }
 \label{fig:3d}
\end{figure}

In Fig.~\ref{fig:3d}, we show DFT results for the solvent density profile between wall and solute
for a solute--solvent size ratio of 20 and solvent--wall distance $h=0$ (i.e. contact between wall and
solute molecular surface; left panel) and $h=\sigma$ (i.e. the end of the depletion region; right panel).
There is strong adsorption at the apex of the annular wedge, given by the coordinates $z=0$ and
$\rp = r_0 = \sqrt{2R(h-\sigma)-(h-\sigma)^2}$, as reflected by the main peak. Along the wall ($z=0$),  we
observe strong structuring which is mainly dictated by packing considerations. 
The apex peak corresponds to a ring of solvent spheres centered at $\rp=r_0$ (for $h=0$) or just one
sphere near $\rp=0$ (for $h=\sigma$).  
The second--highest peak in both panels of Fig.~\ref{fig:3d} appears where the distance between wall
and solute sphere, measured along the $z$--axis, is approximately $2\sigma$ and thus two rings of spheres
fit between solute and wall. However, for $h=0$ (left panel of Fig.~\ref{fig:3d}) packing in
$\rp$--direction leads to further structuring of the density profile close to the apex
of the wedge.
    
\psfrag{rho*}{$\rho^*$}
\psfrag{z/s}{$\frac{z}{\sigma}$}
\psfrag{rp/s}{$\frac{\rp}{\sigma}$}

\begin{figure}
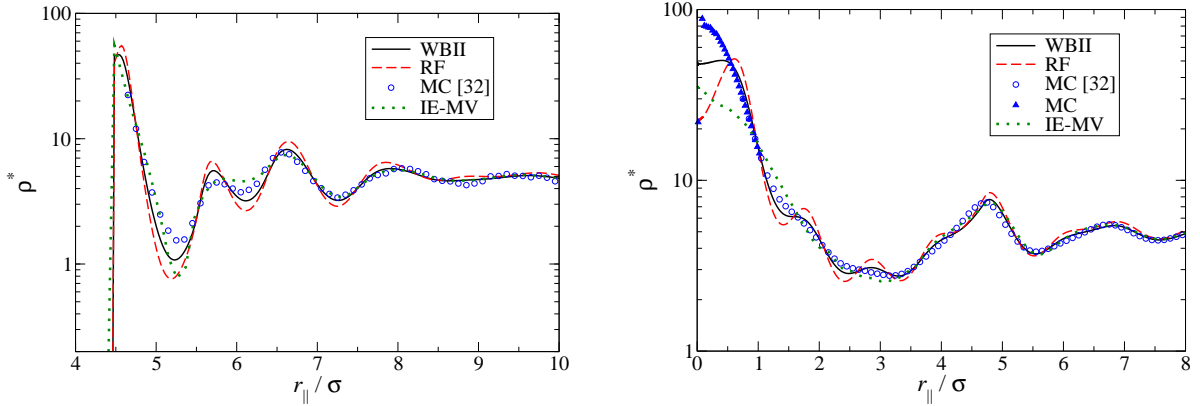

 \begin{center}
   \epsfig{file=fig5a.eps,width=7.5cm} \hspace*{5mm} 
   \epsfig{file=fig5b.eps, width=7.5cm} 
 \end{center}
 \caption{ (color online) Comparison between the various theoretical methods and MC simulations 
(ours and from Ref.~\cite{Her07}) for the density 
at the wall (averaged over a distance of 0.02 $\sigma$). Left panel: solute--wall distance $h=0$,
right panel: $h=\sigma$. The  solute--solvent ratio is 20 and the solvent bulk density
is $\rho_s^*=0.764$. Note the logarithmic scale on the density axis.   }
 \label{fig:rho_wall_20}
\end{figure}

The integral of the solvent density at the wall (i.e. the adsorption at the wall) directly 
determines the depletion force, see Eq.~(\ref{eq:fwdef}). Therefore, we analyze 
the density at the wall further. In Fig.~\ref{fig:rho_wall_20} we show 
$\rho(\rp,z=0)$ computed with FMT--RF, FMT--WBII, IE--MV and compare the results to
the simulation data of Ref.~\cite{Her07}. Overall, DFT performs quite well but it tends to
overestimate the structuring of the profile. The currently most consistent version, FMT--WBII,
improves over FMT--RF particularly in this respect. 
The integral equation results for the different closure are very similar to each other
and are approximately of the same quality as the solutions of FMT--RF. 
Note that the simulation data of Ref.~\cite{Her07} have been averaged    
over a distance of 0.02 $\sigma$ from the wall. This average was also applied to the
DFT data (which are computed with mesh size $\Delta z = 0.002\, \sigma$), whereas for the
IE results we show a suitable weighted average of the zeroth and first bin ($\Delta z = 0.05$),
assuming linear interpolation. As seen in the right panel of Fig.~\ref{fig:rho_wall_20}, IE--MV
gives a much lower density close to the apex of the wedge compared to the other methods. This is
presumably due to the lower resolution in $z$--direction which is possible for IE (see App.~\ref{app:ie}).

\begin{figure}
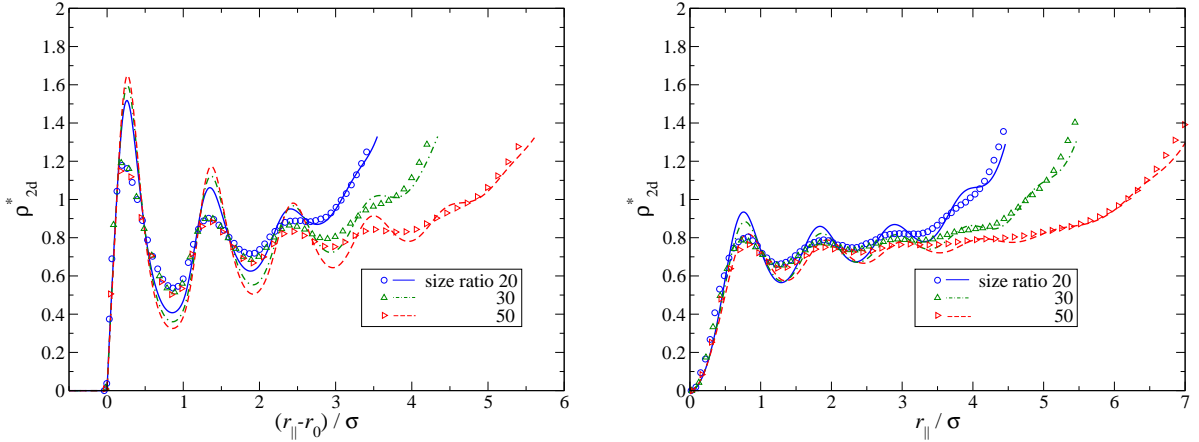

 \begin{center}
   \epsfig{file=fig6a.eps,width=7.5cm} \hspace*{5mm} 
   \epsfig{file=fig6b.eps, width=7.5cm} 
 \end{center}
 \caption{ (color online) Effective two--dimensional solvent density in the annular wedge for various solute--solvent
size ratios. Lines correspond to DFT results (FMT--WBII) and symbols show results of our MC calculations.
 Left panel: solute--wall distance $h=0.95\,\sigma$,
right panel: $h=\sigma$. The solvent bulk density is $\rho_s^*=0.764$.
The curves are plotted up to the point where the width of the annular wedge reaches the value
of $\sigma$.   }
 \label{fig:rho2d}
\end{figure}

To further investigate the issue of the dominant packing mode in the annular wedge, we performed
DFT calculations also for the additonal size ratios 10, 30, 50 and 100 for the same solvent density
$\rho_s^*=0.764$. Packing in $\rp$--direction can be monitored conveniently by viewing the annular
wedge close to the apex as a quasi--2d system (see Fig.~\ref{fig:slit} and  the remarks in the Introduction).  
We define an effective, two--dimensional solvent density by 
\bea
  \rho_{\rm 2d}(\rp) = \int_0^{z_s(\rp)} dz\, \rho(\rp,z) \;,
\eea  
where $z_s(\rp) = R-(\sigma-h) - \sqrt{R^2-\rp^2} \approx \rp^2/(2R) -(\sigma-h)$ defines
the surface of the exclusion sphere around the solute (and thus the radius--dependent width of the annular wedge).
In Fig.~\ref{fig:rho2d} we show results for $\rho_{\rm 2d}(\rp)$ for two configurations with
slightly different solute--wall distances: $h=0.95\,\sigma$ (left panel) and $h=\sigma$ (right panel). 
In the latter case, one solvent sphere fits exactly between solute and wall at the apex, nevertheless
$\rho_{\rm 2d}$ vanishes there due to the vanishing slit width, $z_s(\rp \to 0) \to 0$. Note that
this is also a consequence of an application of the potential distribution theorem \cite{Hen86} which 
states that the 3d density reaches a finite, maximum value of $\exp(\beta\mu^{ex})$ in a planar
slit of vanishing width which is equivalent to a vanishing 2d density ($\mu^{ex}$ is the excess
chemical potential of the bulk solvent coupled to the slit). However, when the slit widens, the
2d density quickly rises \cite{Oet04}. For $h=\sigma$ (right panel in Fig.~\ref{fig:rho2d}), this leads
to the picture of an effective 2d system with a small, soft and repulsive obstacle centered at 
$\rp =0$ which induces moderate layering in the 2d--``bulk" with an effective ``bulk" density
$\rho_{{\rm 2d},s} \approx 0.7 \dots 0.8$. For $h=0.95\,\sigma$ (left panel in Fig.~\ref{fig:rho2d}),
the obstacle in the center has a radial extent of $r_0 \approx \sqrt{R\sigma/10}$ which ranges from   
0.75 $\sigma$ (size ratio 10) to 2.25 $\sigma$ (size ratio 100) and induces stronger layering
in the effective 2d system. The oscillations occur around the same ``bulk" density as in the case
$h=\sigma$. 
The simulation results show weaker oscillations than the DFT results. Since 
the first peak, e.~g., occurs at wedge widths $<0.03\,\sigma$, we are testing the dimensional crossover 
properties of DFT--FMT from 3d to 2d with a sensitive probe. 
The difference between the  results of the simulations and the DFT versions employed here are in line
with previous investigations on the dimensional crossover properties of FMT \cite{Ros97}. There 
it has been found that the strict 2d limit of FMT--RF results in a somewhat peculiar
(integrable) divergence in the hard disk direct correlation function $c_{\rm 2d}(r)$ for $r \to 0$
and overestimated peaks in the corresponding structure factor. The tensor weight modifications
introduced in Ref.~\cite{Tar00} result in better dimensional crossover properties \cite{Tar02}
and might improve the DFT results close to the apex of the annular wedge.  A more detailed
investigation of tensor--weighted FMT with regard to the correlations in narrow planar slits 
and also in the annular wedges considered here is certainly of interest (although the algebra 
of App.~\ref{app:fmt_min} becomes much more extended in this case).      

Our observations should be compared with the phenomenological modelling of Ref.~\cite{Oet04}
(``annular slit approximation"). 
There, the effective 2d system was approximated by an idealized system of hard disks, layering
around a hard cavity of radius $r_0$ centered around $\rp=0$. The bulk density $\rho_{{\rm 2d},s}$ of
this hard disk system was determined via the self--consistency condition   
$p_{\rm 2d}(\rho_{{\rm 2d},s}) = -2\gamma (\rho_s)$ where $p_{\rm 2d}$ is the 2d pressure and $\gamma$
is the surface tension of a hard wall immersed in a bulk solvent of density $\rho_s$. 
Employing scaled particle theory \cite{Hel61}, this yields $\rho_{{\rm 2d},s}^* \approx 0.66 $ for
$\rho_s^* = 0.764$. This value is a bit smaller than the one inferred from the 2d density profiles
of Fig.~\ref{fig:rho2d}. The main difference between the idealized hard disk system of 
Ref.~\cite{Oet04} and the effective 2d system as reflected in Fig.~\ref{fig:rho2d} lies probably
in the softness of the central cavity obstacle and also the softness of the effective particle interactions
due to the widening of the slit. Apart from that the overall picture of Ref.~\cite{Oet04} is confirmed
well. In particular, this implies an important observable consequence: The cavity circumference 
(of approximate length $2\pi r_0$) induces a line contribution $2\pi r_0\,\gamma_{\rm 2d}(\rho_{{\rm 2d},s})$ 
to the insertion free energy  of the solute near the wall. Since the insertion free energy is
equivalent to the depletion potential up to an additive constant, the latter acquires a term
$\propto \sqrt{R(\sigma-h)}$,  the corresponding term in the depletion force is 
$\propto \sqrt{R/(\sigma-h)}$. (Note that for $h\to\sigma$ the cavity radius should stay finite as
reflected in Fig.~\ref{fig:rho2d} (right panel), thus the divergence of the depletion force there
according to the geometrical argument of equating the cavity radius with $r_0$ is unphysical.)    
The interpretation of this line energy term within the more general framework of morphometric
thermodynamics will be given below. 

\begin{figure}
 \begin{center}
   \epsfig{file=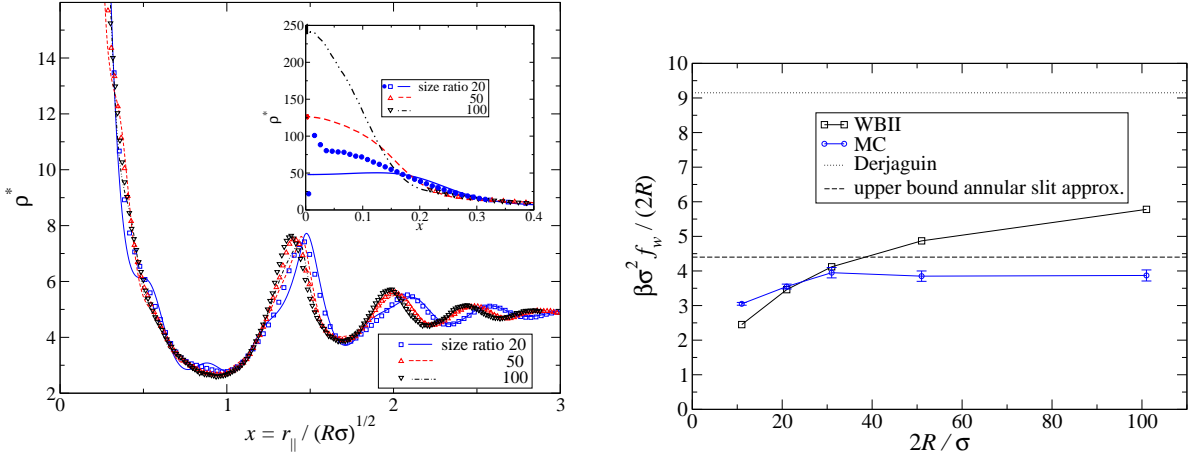,width=7.5cm} \hspace*{5mm} 
   \epsfig{file=fig7b.eps,width=7.5cm} \hspace*{5mm} 
 \end{center}
 \caption{(color online) Left panel: density at the wall for solute--wall distance $h=\sigma$ and the three
solute--solvent size ratios 20, 30 and 100. Lines are FMT--WBII data, symbols
are MC simulation results from Ref.~\cite{Her08}. Right panel: The depletion force between solute and wall
at $h=\sigma$. The solvent bulk density is $\rho_s^*=0.764$.    }
 \label{fig:h1scaling}
\end{figure}

We have seen that in radial direction, the packing of the solvent spheres close to the apex of the  
annular wedge is well understood by the quasi--2d picture developed above. For the particular
point $h=\sigma$, there is only moderate layering in $\rp$--direction, thus the full density profile
should be determined mainly by packing in $z$--direction. Indeed, for $h=\sigma$ we observe a quite remarkable
collapse of the density profiles at the wall when plotted as a function of 
$\hat r = \rp/\sqrt{R\sigma}$, see Fig.~\ref{fig:h1scaling}, indicating the relative unimportance
of packing in radial direction.
According to  Eq.~(\ref{eq:fwdef}), perfect scaling would imply that the depletion force
at $h=\sigma$ is proportional to $R$.
The DFT data violate  scaling for
$\hat r < 0.2$ (i.e. when the slit width is smaller than 0.02 $\sigma$), as can be seen 
from Fig.~\ref{fig:h1scaling} (left panel, inset). This leads to a slow increase of the scaled depletion
force $f_w/(2R)$, see Fig.~\ref{fig:h1scaling} (right panel).
In contrast, the simulation results of Ref.~\cite{Her08} indicate that $f_w/(2R)$ stays constant for large
$R$, quite in agreement with an upper bound derived within the annular slit approximation 
of Ref.~\cite{Oet04} which reads $f_w/(2R) = - 2 \pi \gamma(\rho_s) (1-\eta_{\rm 2d})$ where
$\eta_{\rm 2d} = (\pi/4)\,\rho_{{\rm 2d},s}\sigma^2$ is the 2d ``bulk" packing fraction.
Note that the value of the Derjaguin approximation at $h=\sigma$, $f_w/(2R) = - 2 \pi \gamma(\rho_s)$ 
is completely off the simulation as well as the DFT values (see next subsection). 

\subsubsection{Geometric interpretation of the depletion force}

Morphological (morphometric) thermodynamics \cite{Koe04} provides a powerful interpretation of the depletion interaction,
as has been shown very recently in the special case of the solute--solute interaction \cite{Oet09}.  
Up to a constant, the depletion potential is nothing but the solvation free energy of the wall and the
solute.  
In morphological thermodynamics  the solvation free energy $F_{\mathrm{sol}}$
of a body
is separated into geometric measures defined by its surface. As the surface, we take the
solvent--accessible surface, i.e. the body surface of the combined wall--solute object is defined by the dashed
lines in Fig.~\ref{fig:setup} (a). 
These geometric measures are
the enclosed volume $V$, the surface
area $A$, the integrated mean and Gaussian curvatures $C$ and $X$,
respectively. To each measure there is an associated  thermodynamic coefficient: the
pressure $p$, the planar wall surface tension $\gamma$ and two bending
rigidities $\kappa$ and $\bar \kappa$ such that
\begin{equation} \label{solv}
F_{\mathrm{sol}} = p V + \gamma A + \kappa C + \bar{\kappa} X.
\end{equation}
Due to the separation of the solvation free energy into geometrical measures
and geometry independent thermodynamic coefficients, it is possible to obtain
the coefficients $p$, $\gamma$, $\kappa$ and $\bar{\kappa}$ in simple
geometries. The pressure $p$ is a bulk quantity of the fluid. The surface
tension $\gamma$ accounts for the free energy cost of forming an inhomogeneous
density distribution close to a planar wall. If the wall is curved the
additional free energy cost is measured by $\kappa$ and $\bar{\kappa}$. It is
possible to obtain all four coefficients from a set of solvation free energies
of a spherical particle with varying radius. For a hard-sphere solute in a
hard-sphere solvent very accurate analytic expressions for the thermodynamic
coefficients are known from FMT--WBII \cite{Han06}, see also App.~\ref{app:measure_coefficients}. 

\begin{figure}
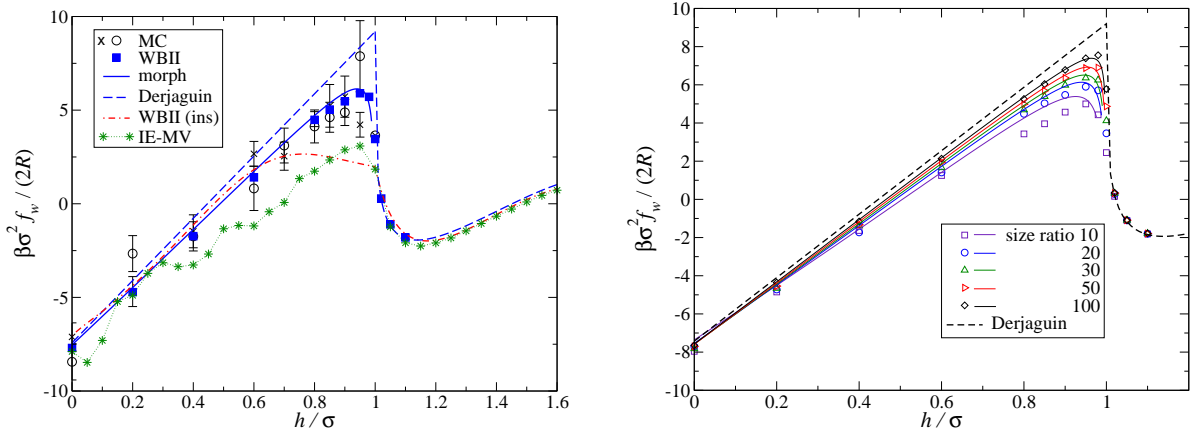

 \begin{center}
   \epsfig{file=fig8a.eps,width=7.5cm} \hspace*{5mm} 
   \epsfig{file=fig8b.eps,width=7.5cm} 
 \end{center}
 \caption{(color online) Depletion force between wall and solute for solvent bulk density $\rho_s^*=0.764$.
Left panel: Solute--solvent size ratio 20. Simulation data  are shown 
by crosses ($\times$) and circles ({\tt o}) and differ by a slightly different choice
of $\rho_w$ in Eq.~(\ref{eq:fwdef}) (see Refs.~\cite{Her06,Her07} for details). Results from
explicit FMT--WBII minimization are given by the full squares and results from IE--MV are shown by asterisks. The dash--dotted line shows the insertion route 
result using the WBII functional and the dashed line gives the Derjaguin approximation. 
The excess surface free energy of two
parallel hard walls, needed for the Derjaguin approximation, has been calculated with the
WBII functional. Right panel: depletion force for different solute--solvent  size ratios. Symbols
are results from explicit FMT--WBII minimization, lines in ascending order are morphometric results
for size ratios 10, 20, 30, 50, 100 and $\infty$ (Derjaguin approximation). }
 \label{fig:hforce}
\end{figure}

For the particular case considered here, the depletion potential $W(h)$ follows by subtracting
the sum of the solvation free energies of a single wall and a single solute sphere. Thus
we find
\bea
 \label{eq:Wdef}
  W(h<\sigma) &=& - p\Delta V - \gamma \Delta A - \kappa \Delta C -4\pi \bar\kappa\;
\eea 
where $\Delta V$ and $\Delta A$ are the volume and surface area of the overlap of the exclusion (depletion)
zones around wall and solute, respectively, and $\Delta C$ is the integrated mean curvature of
that overlap volume. Note that the fourth characteristic, the integrated Gaussian curvature is the 
Euler characteristic which is $4\pi$ for wall and solute at $h\le \sigma$ (one connected body) and
$8\pi$ for wall and solute at $h > \sigma$ (two disconnected bodies). 
This explains the  last term in Eq.~(\ref{eq:Wdef}). The geometric measures $\Delta V$, $\Delta A$  and
$\Delta C$ are given by
\bea
 \Delta V &=& \frac{\pi}{3}(\sigma-h)^2\,(3R-(\sigma-h)) \nonumber \\
         & \stackrel{R\to\infty}{\approx} & \pi R (\sigma-h)^2 + O(R^0) \;, \nonumber\\
 \Delta A &=& 4\pi R(\sigma-h) - \pi(\sigma-h)^2 \nonumber\\
       & \stackrel{R\to\infty}{\approx} & 4\pi R(\sigma-h) + O(R^0) \;, \\
 \Delta C &=& 2\pi(\sigma-h) + \pi\sqrt{2R(\sigma-h)-(\sigma-h)^2}
    \left( \frac{\pi}{2} + \arcsin \left( 1- \frac{\sigma-h}{R}\right) \right) \nonumber\\ 
  &   \stackrel{R\to\infty}{\approx} & \pi^2 \sqrt{2R(\sigma-h)} + O(R^0)  \;. \nonumber
\eea
The depletion force follows as $f_w(h) = -\partial W/\partial h$. In the limit $R\to \infty$ it is
given by:
\bea
 \label{eq:forcemorph}
  \frac{f_w(h<\sigma)}{2\pi R} & \stackrel{R\to\infty}{\approx} &  
      p(h-\sigma) - 2\gamma - \kappa\, \frac{\pi}{2} \sqrt{\frac{1}{2R(\sigma-h)}} \;.
\eea
The contribution from the 
integrated mean curvature is subleading in $1/R$ but it is quantitatively important since it induces 
a singularity as $h \to\sigma$ which is 
proportional to $1/(\sigma-h)^{1/2}$. 
This contribution precisely corresponds to the line tension term associated with the circular
apex (of length $2\pi r_0$) of the annular wedge identified in the quasi--2d analysis of Ref.~\cite{Oet04} and
introduced in the last subsection. The line tension in the quasi--2d picture was shown to be
$\gamma_{\rm 2d}(\rho_{{\rm 2d},s})$ and in the geometric analysis it is $-\pi\kappa/2$. There is good
agreement between both expressions \cite{Oet09}.   
Strictly, the singularity associated with this line term is unphysical of course. 
Indeed, it can be shown that the 
morphometric analysis becomes ambiguous very close to $h=\sigma$. One would require that the morphometric 
result for the force is unchanged if another, slightly displaced surface around the solute sphere is
chosen for the geometric analysis. In Ref.~\cite{Oet09}, such an analysis was carried out for the
sphere--sphere geometry and the independence of the force on the chosen surface was demonstrated. However,
if one chooses e.g. the molecular surface (the surface of the set of points which is never covered
by a small solvent sphere), then this surface becomes self--overlapping close to $h=\sigma$ and one
could not expect to ascribe physical significance to surface tensions and mean curvature coefficients
of such overlapping surfaces. For the sphere--wall geometry a similar analysis holds. Note that
the singularity at $h=\sigma$ corresponds to a vanishing circumference of the apex ($r_0=0$), 
however, the 2d analysis of the density distributions at this point yielded a small but finite value for the
``effective" apex radius $r_0$ (see Fig.~\ref{fig:rho2d} (right panel)).

In Fig.~\ref{fig:hforce} (left panel) we show FMT--WBII and IE--MV results for the depletion force
$f_w$ between a big sphere and a wall (size ratio 20,
solvent bulk density $\rho_s^*=0.764$) and compare them to MC results from Refs.~\cite{Her06,Her07}  
as well as to the morphometric analysis. The force according to morphometry is plotted only until
self--intersection of the molecular surface starts. 
The FMT data are described very well by the morphometric results;
the sharp drop in the depletion force close to $h=\sigma$ appears to mimick the behavior   
of the singular term $\propto \partial \Delta C/\partial h$.
The MC data suffer from relatively large error bars (except for the point $h=\sigma$ but are overall
consistent with both the FMT data and the morphometric analysis.
The scatter in the IE--MV data is due to the comparatively low resolution in $z$--direction as compared
to the DFT data ($\Delta z = 0.05\,\sigma$ vs. $\Delta z = 0.002\,\sigma$). As before, all 
considered IE closures give
similar results, but these correspond to an overall more attractive force compared to DFT and MC.
 Two other approximations for the depletion force are
shown in  Fig.~\ref{fig:hforce} (left panel): the DFT insertion route and the Derjaguin approximation. Both
approximations fail to describe the depletion force near $h=\sigma$. 
In the insertion route to DFT (see Ref.~\cite{Rot00} for its derivation),
one circumvents the explicit calculation of the density profile around the fixed wall and solute
by exploiting the relation:
\bea
 \label{eq:ins}
  \beta W(\vect x) &=& \lim_{\rho_b\to 0} 
   \left. \frac{\delta {\cal F}^{\rm ex}}{\delta \rho_b(\vect x)}  \right|_{\rho_s(\vect r) = 
  \rho_w(z)} - \mu^{\rm ex}_b(\rho_s) \;.
\eea
Here, $\mu^{\rm ex}_b(\rho_s)$ is the excess chemical potential for inserting one solute into the
solvent with density $\rho_s$. The functional derivative of the {\em mixture} functional has to be
evaluated in the dilute limit for the solutes, $\rho_b\to 0$, and with the solvent density profile given 
by the profile at a hard wall ($\rho_w(z)$). This method relies on an accurate representation
 of mixture effects for large size ratios in the free energy functional.   
Therefore, the observed deviations in the depletion force presumably follow
from the insufficient representation of higher-order direct
correlation functions of strongly asymmetric mixtures in
the present forms of FMT \cite{Cue02}. The Derjaguin approximation, on the other hand, corresponds to the
leading term for $R\to \infty$ in Eq.~(\ref{eq:forcemorph}) inside the depletion region
($h<\sigma$), but is easily extended to $h \ge \sigma$ \cite{Hen02,Oet04}:
\bea
 \label{eq:fw_der}
  \frac{f^{\rm Derjaguin}_w}{2\pi R} & = & \left\{ \begin{array}{lll}  
      p(h-\sigma) - 2\gamma & & (h<\sigma) \\
      \gamma_{\rm slit}(h)  - 2\gamma  & & (h\ge \sigma) \end{array}\right. \;,  
\eea 
with $\gamma_{\rm slit}(h)$ denoting the excess surface energy of two parallel hard walls 
at distance $h$ forming a slit. According to Eq.~(\ref{eq:fwdef}), the Derjaguin force
scales with $R$ (``colloidal limit") which is an excellent approximation outside the depletion
region. (Incidentally, outside the depletion region all methods and the Derjaguin approximation
agree with each other.) Inside the depletion region, the neglected mean curvature (or line tension) term is very significant.

Explicit FMT--WBII results for the depletion force for size ratios up to 100 are shown in 
Fig.~\ref{fig:hforce} (right panel) and compared to the morphometric analysis (symbols vs. full lines).
Whereas for size ratio 10 some discrepancies are still visible, they have more or less vanished for 
size ratio 100 (except for the point $h=\sigma$, see the previous discussion). Thus the morphometric
form for the depletion force (Eq.~(\ref{eq:forcemorph})) can be regarded as the appropriate ``colloidal
limit".  

\subsection{Solute--solute interaction}

\label{sec:solute_solute}

We now turn to the case of the depletion force between two big solute particles which are immersed
in the hard solvent. Considering the evidence from the sphere--wall case, we would expect that
for large size ratios, the morphometric picture reliably 
describes the depletion force. Since the geometry of the overlap volume between two spheres is
different from that between a sphere and a wall, we find also a slightly different form for the morphometric
force in the limit $R\to\infty$:
\bea
 \label{eq:force_s_morph}
  \frac{f(h<\sigma)}{\pi R} & \stackrel{R\to\infty}{\approx} &  
      p(h-\sigma) - 2\gamma - \kappa\, \frac{\pi}{2} \sqrt{\frac{1}{R(\sigma-h)}} \;.
\eea
Thereby, one sees that the scaling relation $f_w = 2f$, valid within
the Derjaguin approximation, is violated through the appearance of the mean curvature (line tension) term. 
For moderate solvent densities up to $\rho_s^*=0.7$  the morphometric form 
(\ref{eq:force_s_morph}) gives a good description of the depletion force for size ratios 
$\alpha \agt 10$. This has been reported in Ref.~\cite{Oet09}. An analysis of the full solvent density
profiles between the two solutes reveals the same features as described in Sec.~\ref{sec:wall_solute}. 
  
\begin{figure}
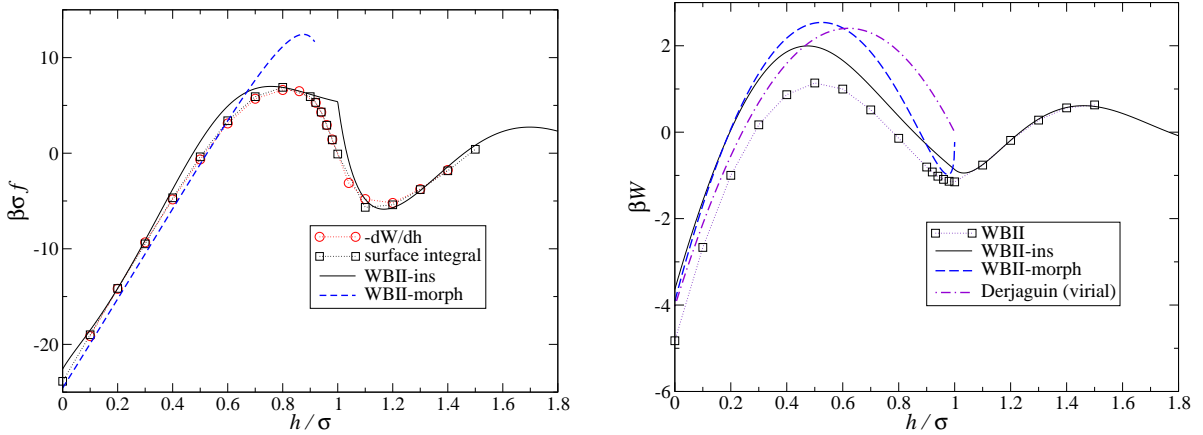

 \begin{center}
   \epsfig{file=fig9a.eps,width=7.5cm} \hspace*{5mm} 
   \epsfig{file=fig9b.eps,width=7.5cm} 
 \end{center}
 \caption{(color online) Left panel: Depletion force between two solutes. Circles correspond to
the centered difference of the FMT-WBII data for the depletion potential (right panel), while
squares give the FMT--WBII depletion force according to the surface integral in Eq.~(\ref{eq:fdef}). Right panel: Depletion 
potential between two solutes. The curve for the morphometric potential has been shifted by -1.  
The solvent density is $\rho_s^*=0.8$ and the solute--solvent size ratio is 5.  }
 \label{fig:f_a5r0.8}
\end{figure}

\begin{figure}
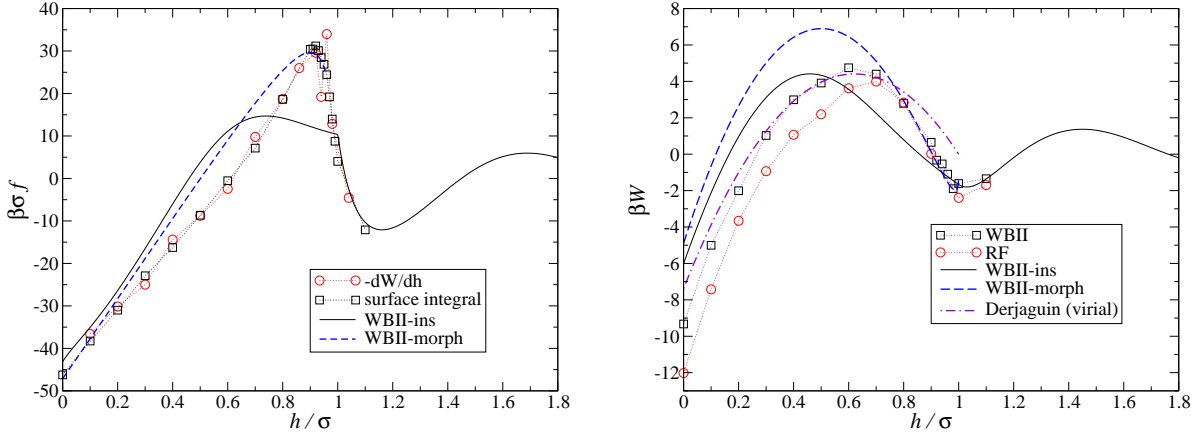

 \begin{center}
   \epsfig{file=fig10a.eps,width=7.5cm} \hspace*{5mm} 
   \epsfig{file=fig10b.eps,width=7.5cm} 
 \end{center}
 \caption{(color online) Left panel: Depletion force between two solutes. Circles correspond to
the centered difference of the FMT-WBII data for the depletion potential (right panel), while
squares give the FMT--WBII depletion force according to the surface integral in Eq.~(\ref{eq:fdef}). Right panel: Depletion 
potential between two solutes. Squares correspond to FMT--WBII results, and circles to FMT--RF results. 
The curve for the morphometric potential has been shifted by -2.  
The solvent density is $\rho_s^*=0.8$ and the solute--solvent size ratio is 10.  }
 \label{fig:f_a10r0.8}
\end{figure}

It is interesting to investigate the regime where the morphometric description is expected to fail.
This will happen when there are strong correlations in the whole annular wedge, i.e. where packing in both
the $\rp$-- and the $z$--direction becomes important. Typically, intermediate size ratios and higher solvent  
densities will induce these strong correlations. Therefore, we have investigated the depletion
force for size ratios 5 and 10 and solvent densities $\rho_s^*=0.8$ and $\rho_s^*=0.9$. 
In Figs.~\ref{fig:f_a5r0.8} and \ref{fig:f_a10r0.8} we show results for the depletion force and potential
at $\rho_s^*=0.8$ for size ratios 5 and 10 respectively. For size ratio 5, the explicit DFT calculations 
yield a depletion force which does not correspond to the morphometric result at all. It is closer to
the insertion route (see (Eq.~\ref{eq:ins})), though systematically lower. This yields a depletion potential
(Fig.~\ref{fig:f_a5r0.8} (right panel)) which is about 1 $k_{\rm B}T$ or 25\% more attractive at contact
than that of the insertion route. For size ratio 10, the explicit DFT data for the depletion force 
reflect the morphometric form as $h\to \sigma$, i.e. they give again evidence for 
importance of the mean curvature (line tension) term in Eq.~(\ref{eq:force_s_morph}). However, 
away from that regime, the depletion force deviates significantly from both the morphometric form
and the insertion route result such that the depletion potential at contact 
(Fig.~\ref{fig:f_a10r0.8} (right panel)) is about 3 $k_{\rm B}T$ or 50\% more attractive at contact
than that of the insertion route (for FMT--WBII). 

Note that at such a high solvent density ($\rho_s^*=0.8$) there are already significant deviations
between FMT--RF and FMT--WBII (Fig.~\ref{fig:f_a10r0.8} (right panel), circles and squares). 
Since FMT--WBII has been designed to improve thermodynamic and morphometric consistency at higher 
densities, the corresponding results can be assumed to be more trustworthy. As an additional check of
the numerics, we calculated the depletion force in two ways: (a) via the surface integral in
Eq.~(\ref{eq:fdef}) and (b) via the centered difference of the results for the depletion potential 
(see Figs.~\ref{fig:f_a5r0.8} and \ref{fig:f_a10r0.8} (left panel), squares and circles). The latter is 
simply obtained as
\bea
  W(h) &=& \left.\Omega [\rho(\vect r); h]\right|_{\rho(\vect r) = \req(\rp,z;h)} -
  \left.\Omega [\rho(\vect r); h\to \infty]\right|_{\rho(\vect r) = \req(\rp,z;h\to\infty)} \;, 
\eea 
where the grand potential $\Omega$ and the equilibrium density $\req(\rp,z;h)$ around the two solutes
at distance $h$ are determined by the basic DFT equations (\ref{eq:omegadef}) and (\ref{eq:rhoeq_main}),
respectively. The agreement between routes (a) and (b) is very good, save for some ``fluctuations"
in route (b) close to the point $h=\sigma$ where the density oscillations in the wedge are most pronounced. 
The equality of both routes checks a  sum rule check similar to the hard wall sum rule for the problem
of one wall immersed in solvent. Weighted--density DFT's usually fulfill the latter \cite{Swo89}. 

\begin{figure}
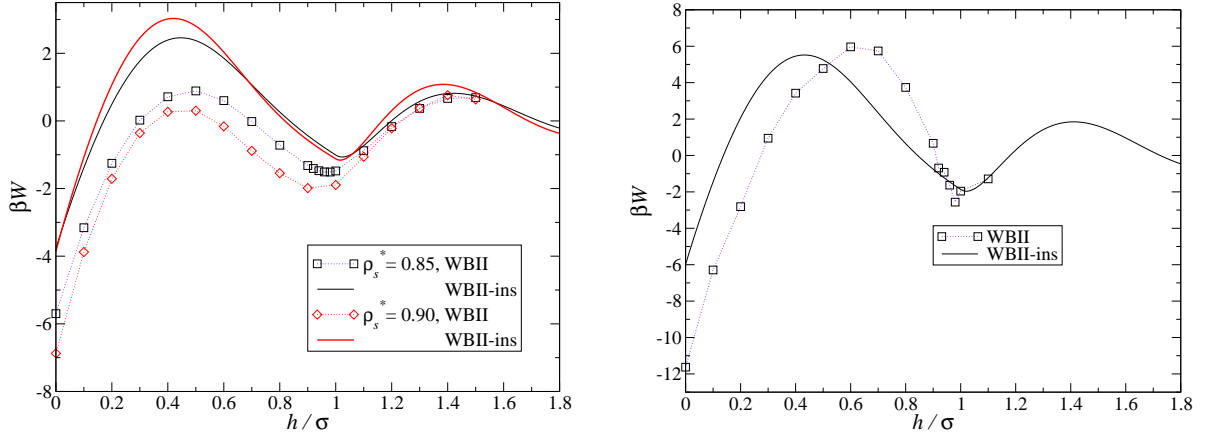

 \begin{center}
   \epsfig{file=fig11a.eps,width=7.5cm} \hspace*{5mm} 
   \epsfig{file=fig11b.eps,width=7.5cm} 
 \end{center}
 \caption{(color online) Depletion potential between two solutes, comparison
between explicit FMT--WBII data and the insertion route. Left panel: 
Solute--solvent size ratio 5 and solvent densities $\rho_s^*=0.85$ and 0.9.
Right panel: Size ratio 10 and $\rho_s^*=0.85$.
}
 \label{fig:w_r0.85_0.9}
\end{figure}

The differences between the insertion route and the explicit FMT data becomes more
dramatic for higher  densities.
In Fig.~\ref{fig:w_r0.85_0.9} we show results for the depletion potential
for a solute--solvent size ratio 5 at solvent densities 0.85 and 0.9 (left panel)
and for a size ratio 10 at density 0.85 (right panel). For size ratio 5 the repulsive barrier
has almost vanished for $\rho_s^*=0.9$, and the potential well at contact is almost twice as
deep as compared with the insertion route. For size ratio 10, the barrier remains (albeit at a 
different location) but the well depth is equally enhanced by a factor 2 
compared with the insertion route as for size ratio 5.

The significant deviations of the depletion potential (according to the explicit FMT--WBII results) from
the previously known approximations (Derjaguin approximation and insertion route) sheds some doubts on
the quantitative accuracy of simulations of binary hard sphere mixtures (with asymmetries between 5 and 10),
employing an effective one--component approach for the solutes interacting with their depletion potential.
In Ref.~\cite{Dij99} the phase diagram of the mixture was scanned in that way using a virial expansion
to third order in the Derjaguin approximation for the depletion potential (see the dot--dashed curves in
Figs.~\ref{fig:f_a5r0.8} and  \ref{fig:f_a10r0.8} (right panel)). Especially for size ratio 5, the explicit DFT
results predict a depletion potential with much smaller barrier and a deeper well at contact. This might affect 
the phase diagram in the corner where the packing fraction of the solutes is low and
the solvent packing fraction is high. In Ref.~\cite{Amo08} gellation in  hard sphere--like, asymmetric mixtures 
is investigated with an integral equation approach (reference functional approximation \cite{Aya09}) to the
depletion potential, akin to the insertion route. We expect quantitative deviations at higher solvent 
densities compared to explicit DFT calculations which again affect the corner of the phase diagram
with low solute and high solvent packing fraction.

Finally we want to mention that the explicit DFT methods introduced here, together with the morphometric
form of the depletion force in Eq.~(\ref{eq:force_s_morph}), can be used to improve available analytic forms 
\cite{DHen08,DHen09} for the contact values of the pair correlation functions in hard--sphere mixtures 
(note that the solute--solute contact value is given by $g_{bb}(2R_b) = \exp[-\beta W(0)]$).

\section{Summary and conclusions}

\label{sec:summary}

We have presented a detailed analysis of the relationship between depletion forces 
(solute--wall and solute--solute) in a hard solvent and the associated density distributions
around the fixed wall/solute particles, using mainly density functional theory of fundamental measure type
 complemented with Monte--Carlo simulations and integral equation techniques. For large solute--solvent
size ratios, the depletion force is strongly linked to the quasi two--dimensional confinement between 
the solutes (or the solute and the wall). We have shown that the properties of this quasi 
two--dimensional confined system are reflected in the depletion force by a line--tension term
which in turn can be obtained quite generally through morphological thermodynamics. Thus we can formulate
an appropriate ``colloidal" limit for the depletion force in hard--sphere mixtures:
Eq.~(\ref{eq:forcemorph}) for the solute--wall case and Eq.~(\ref{eq:force_s_morph}) for the solute--solute case. 
This new formulation improves significantly over the Derjaguin approximation which
is a frequently employed tool to estimate colloidal interactions in many circumstances.

Although our analysis has been restricted to hard spheres only, the formulation of morphological
thermodynamics is very general such that it can be expected that the colloidal limit for the depletion 
force holds also in mixtures with more general interparticle interactions. However, more explicit
tests in this direction are necessary.

For intermediate size ratios 5 and 10 and higher solvent densities $\rho_s^* \ge 0.8$ we found strong
deviations in the solute--solute depletion force between the explicit density functional results on the one
hand and the Derjaguin approximation or morphometry on the other hand. This is a result of the strong solvent 
correlations in the annular wedge between the solutes. The depletion well at solute contact is significantly more attractive (by almost a factor of two)
than previously estimated. This might have consequences for the phase diagram
of binary hard spheres at low solute and high solvent packing fractions. With recently developed 
techniques \cite{Mil04,Sch09}, this question appears to be tractable in direct simulations.

{\bf Acknowledgment:} V.B., F.P. and M.O. thank the German Science Foundation for financial
support through the Collaborative Research Centre (SFB--TR6) ``Colloids in
External Fields", {project N01}. T.S. thanks the German Science Foundation for financial
support through an Emmy Noether grant.

\begin{appendix}

\section{Fundamental measure functionals for hard spheres and associated thermodynamic coefficients} 

\label{app:measure_coefficients}

We consider fundamental measure functionals involving no tensor--weighted densities
which are defined by the following functional for the excess free energy of a
hard sphere mixture of solvent and solute particles, described by the
 density distribution $\rho(\vect r)=\{\rho_s(\vect r), \rho_b(\vect r) \}$:
\bea
 \label{eq:fhs}
 \beta {\cal F}^{\rm ex} &=& \int d\vect r \, \Phi( \{\vect n[\rho (\vect r)]\} ) \;,  \\
   \Phi( \{\vect n[\rho (\vect r)]\} ) &=&   -n_0\,\ln(1-n_3) + 
      \varphi_1(n_3)\;\frac{n_1 n_2-\vect n_1 \cdot \vect n_2}{1-n_3} + 
      \varphi_2(n_3)\;\frac{n_2^3-3n_2\, \vect n_2\cdot \vect n_2}{24\pi(1-n_3)^2}\;. 
   \nonumber
\eea
Here, $\Phi$, is  a 
free energy density which is a function of a set of weighted densities
$\{\vect n (\vect r)\} = \{ n_0,n_1,n_2,n_3,\vect n_1,\vect n_2\}$
with four scalar and two vector densities. These are related to the
density profile $\rho(\vect r)$ by
\bea
 \label{eq:wddef}
  n_\alpha =  \sum_i \int d\vect r' \rho_i(\vect r') \, w^\alpha_i(\vect r- \vect r')
           = \sum_i \rho_i * w^\alpha_i \; ,
\eea
and the  hereby introduced  weight functions,
$\{\vect w_i(\vect r)\} = \{w_i^0,w_i^1,w_i^2,w_i^3, \vect w_i^1,\vect w_i^2\}$,
depend on the hard sphere radii $R_i=\{R_s,R_b\}$ of the solvent and solute
particles as follows:
\bea
 w^3_i = \theta(R_i-|\vect r|)\;, \qquad
 w^2_i = \delta(R_i-|\vect r|)\;, \qquad w^1_i = \frac{w^2_i} {4\pi R_i}\;, 
  \qquad w^0_i = \frac{w^2_i}{4\pi R_i^2}\;,
  \nonumber\\
 \vect w^2_i =\frac{\vect r}{|\vect r|}\delta(R_i-|\vect r|)\;,
 \qquad \vect w^1_i = \frac{\vect w^2_i}{4\pi R_i} \;.
\eea
The excess free energy functional in Eq.~(\ref{eq:fhs}) is completed upon specification
of the functions $\varphi_1(n_3)$ and $\varphi_2(n_3)$. With the choice
\bea
 \label{eq:frf}
 \varphi_1 = 1\;, \qquad \varphi_2 = 1
\eea
we obtain the original Rosenfeld functional \cite{Ros89}. Upon setting
\bea
 \label{eq:fwb}
 \varphi_1 &= & 1 \\
 \varphi_2 & = & 1 - \frac{-2n_3+3n_3^2 - 2(1-n_3 )^2 \ln(1-n_3 ) }
                          {3 n_3^2}  \nonumber    
\eea
we obtain the White Bear functional \cite{Rot02,Yu02}, consistent with the quasi--exact
Carnahan--Starling equation of state. Finally, with
\bea
 \label{eq:fwbII}
 \varphi_1 & = & 1 + \frac{2n_3-n_3^2 + 2(1-n_3) \ln(1-n_3 )}{3n_3} \\ 
 \varphi_2 & = & 1 - \frac{2n_3-3n_3^2 + 2n_3^3 + 2(1-n_3 )^2 \ln(1-n_3 ) }
                          {3 n_3^2}  \nonumber    
\eea
the recently introduced White Bear II functional is recovered.  

Next we briefly recapitulate the determination of the thermodynamic coefficients $p$ (pressure),
$\gamma$ (hard wall surface tension), $\kappa$  and $\bar\kappa$ (bending rigidity of the
integrated mean and Gaussian curvature, respectively) \cite{Han06}. Consider the free energy of 
insertion $F_b$ for a big solute particle of radius $R_b$ and associated radius
$R=R_b+R_s$  of the exclusion sphere around it. According to Eq.~(\ref{solv}), it is given by:
\bea
\label{eq:fb_sphere}
  F_b = p\, \frac{4\pi}{3}R^3 + \gamma\, 4\pi R^2 + \kappa\, 4\pi R + \bar\kappa \,4\pi \;. 
\eea  
On the other hand, $F_b$ is equivalent to the excess chemical potential of the solutes in a mixture
with solvent particles at infinite dilution ($\rho_b \to 0$). This excess chemical potential is obtained as the
density derivative $\partial /\partial \rho_b$ of the mixture free energy (Eq.~(\ref{eq:fhs})), therefore
we obtain: 
\bea
 \beta F_b  &=& \lim_{\rho_b \to 0} \beta \frac{\partial {\cal F}^{\rm ex}}{\partial \rho_b} \nonumber  \\
\label{eq:fb_func}
      & =& \frac{\partial \Phi}{\partial n_3}\, \frac{4\pi}{3}R_b^3 +
           \frac{\partial \Phi}{\partial n_2}\, 4\pi R_b^2 +
           \frac{\partial \Phi}{\partial n_1}\,  R_b + \frac{\partial \Phi}{\partial n_0}  \:.
\eea
By equating the two expressions for $F_b$ in eqs.~(\ref{eq:fb_sphere}) and (\ref{eq:fb_func}), one
finds the explicit expressions for the thermodynamic coefficients as linear combinations    
of the partial derivatives $\partial\Phi/\partial n_i$. For the Rosenfeld functional, these
coefficients are given by:
\bea
  \beta p &=& \rho_s\, \frac{1+\eta_s+\eta_s^2}{(1-\eta_s)^3} \;, \nonumber\\
  \beta \gamma &=& - \frac{3}{4}\rho_s\sigma \, \frac{\eta_s(1+\eta_s)}{(1-\eta_s)^3} \;,\nonumber\\
  \beta \kappa &=& \frac{3}{4}\rho_s\sigma^2 \, \frac{\eta_s^2}{(1-\eta_s)^3} \;,\\
  \beta \bar \kappa &=& \rho_s\sigma^3 \left( \frac{-2+7\eta_s-11\eta_s^2}{48(1-\eta_s)^3} - 
       \frac{\ln(1-\eta_s)}{24\eta_s} \right)\;.\nonumber
\eea
For the White Bear II functional, the thermodynamic coefficients have been given already  in 
Ref.~\cite{Han06} and  are reproduced here for completeness:
\bea
  \beta p &=& \rho_s\, \frac{1+\eta_s+\eta_s^2-\eta_s^3}{(1-\eta_s)^3} \;,\nonumber\\
  \beta \gamma &=& - \rho_s\sigma \left( \frac{1+2\eta_s+8\eta_s^2-5\eta_s^3}{6(1-\eta_s)^3} +
     \frac{\ln(1-\eta_s)}{6\eta_s} \right) \;,\nonumber\\
\label{eq:coefwbII}
  \beta \kappa &=& \rho_s\sigma^2 \left( \frac{2-5\eta_s+10\eta_s^2-4\eta_s^3}{6(1-\eta_s)^3} +
    \frac{\ln(1-\eta_s)}{3\eta_s} \right)\;,\\
  \beta \bar \kappa &=& \rho_s\sigma^3 \left( \frac{-4+11\eta_s-13\eta_s^2+4\eta_s^3}{24(1-\eta_s)^3} - 
       \frac{\ln(1-\eta_s)}{6\eta_s} \right)\;.\nonumber
\eea
The coefficients of the White Bear II functional are remarkably consistent with respect to
an explicit minimization of the functional around solutes of varying radius and fitting the
corresponding insertion free energy to Eq.~(\ref{eq:fb_sphere}) \cite{Han06}.

\section{Minimization of fundamental measure functionals in cylindrical coordinates}

\label{app:fmt_min}

In the following we consider a one--component fluid of solvent hard spheres only 
($\rho_s(\vect r) \equiv \rho(\vect r)$).
The equilibrium density profile $\req(\vect r)$ of the hard sphere fluid with
chemical potential $\mu=\beta^{-1}\ln(\rho_s\, \Lambda^3) + \mex$
(corresponding to the bulk density $\rho_s$)
in the presence of an arbitrary external potential $V(\vect r)$
is found by minimizing the grand potential
\bea
  \Omega[\rho] &=& {\cal F}^{\rm id}[\rho] +  {\cal F}^{\rm ex}[\rho] - \int
   d\vect r(\mu - V^{\rm ext}(\vect r)) \;,
\eea 
which leads to
\bea
 \label{eq:rhoeq}
  \beta^{-1}\ln \frac{\req(\vect r)}{\rho_s} =-\mu [\req(\vect r)] + \mex - V^{\rm ext}(\vect r) \;. 
\eea
The functional $\mu [\rho(\vect r)]$ is given by
\bea
 \mu [\rho(\vect r)] &=&  \frac{\delta {\cal F}^{\rm ex}[\rho]}
      {\delta \rho(\vect r)} \\
\label{eq:mufunc}
    &=& \beta^{-1} \sum_\alpha \int d\vect r'\, \frac{\partial \Phi}{\partial n_\alpha(\vect r')}
         \,w^\alpha(\vect r'-\vect r) \;.
\eea
For the physical problems of this work, sphere--sphere and wall--sphere geometry, 
the external potential $V^{\rm ext}$ possesses rotational symmetry around the $z$--axis. 
Therefore we work in cylindrical coordinates $\vect r = (\rp\cos\phi,\rp\sin\phi,z)$,
in which the external potential and the density profile depend only on $\rp$ and $z$,
$V\equiv V(\rp,z)$ and $\req \equiv \req(\rp,z)$.  
Eqs.~(\ref{eq:rhoeq}) and (\ref{eq:mufunc}) can be solved with a standard Picard 
iteration procedure or a speed--enhanced scheme (see below). 
The technical difficulty lies in the fast and efficient numerical evaluation of
the weighted densities $n_\alpha=\rho * w^\alpha$ and the convolutions
$\partial \Phi/\partial n_\alpha * w^\alpha$ appearing in Eq.~(\ref{eq:mufunc}).

Convolution integrals are calculated most conveniently in Fourier space where
they reduce to simple products. Generically, the (three--dimensional, 3d) 
Fourier transforms of the weighted densities involve a one--dimensional (1d)
Fourier transform in the $z$--coordinate and a Hankel transform of zeroth or
first order in the $\rp$--coordinate (see Sec.~\ref{sec:wd}). Both the 1d
Fourier transform and the Hankel transform can be calculated using fast Fourier
techniques (see Sec.~\ref{sec:fht}).   

\subsection{Weighted densities}

\label{sec:wd}

The convolution integral, defining the weighted densities in Eq.~(\ref{eq:wddef}), is
calculated by:
\bea
  \vect n (\vect r) = \int \frac{d\vect q}{(2\pi)^3} {\exp(-i\vect q \cdot \vect r)}
   \, \tilde \rho(\qp,q_z)\,\tilde {\vect  w}(\vect q)\;.
\eea
The 3d Fourier transform of the density profile $\rho(\rp,z)$ appearing in
the above equation is given by:
\bea
  \tilde \rho(\qp,q_z) &=& \int d\vect r \exp(i\vect q \cdot \vect r) \rho(\rp,z) \;\\
    &=&  \int_{-\infty}^\infty dz\, \exp(iq_z z)\,\int_0^\infty 2\pi\rp d\rp\,
    J_0(\qp \rp) \,\rho(\rp,z)  \;, \\
    &=& \ft \bt 0 \rho(\rp,z)\;.
\eea
Here, we have introduced shorthand notations $\ft$ for the Fourier transform in the 
$z$--coordinate and $\bt i$ for the Hankel transform in the $\rp$--coordinate
involving the kernel $J_i(\qp\rp)$. A vector in real space is given by
$\vect r = \rp \vect e_{||} + z \vect e_z$, whereas a vector in Fourier space is given by
$\vect q = \qp \vect e'_{||} + q_z \vect e_z$, with $\vect e_{||} \cdot \vect e'_{||} 
= \cos\phi_q$.
The 3d Fourier transforms $\tilde {\vect  w}$  of the set of weight functions 
$\vect w$ reduce to sine transforms due
to radial symmetry and are explicitly given by
\bea
  \tilde w^3(q) & = &  \frac{4\pi R}{q^2} \left( \frac{\sin(qR)}{qR} - \cos(qR)\right)
   \;, \\
  \tilde w^2(q) &=& \frac{4\pi R}{q}\, \sin(qR) \;, \\
   \tilde{\vect  w}^2(\vect q) &=& -i\vect q\, \tilde w^3(q)  \\
\eea
(The remaining weighted densities differ only by a multiplicative factor.)
It is convenient to introduce the following parallel and perpendicular
components of the Fourier transformed vector weights ($k=1,2$):
\bea
   \tilde w^k_{||} &=& i \vect e'_{||} \cdot \tilde{\vect  w}^k(\vect q) \;, \\  
   \tilde w^k_{z} &=& i \vect e_{z} \cdot \tilde{\vect  w}^k(\vect q) \;.
\eea 
Using these definitions, the scalar weighted densities are given by 
\bea
  n_k (\rp, z) = \ft^{-1}\bt 0^{-1}\left[ \tilde \rho(\qp,q_z)\;
    \tilde w^k\left(\sqrt{\qp^2+q_z^2}\right) \right] \;. \qquad
   (k=0\dots 3)\;.
\eea
The vector weighted densities, on the other hand, are given by two components
($k=1,2$):
\bea
 \nonumber
  \vect n_k(\vect r) &=&  n_{k,||}(\rp,z) \vect e_{||} +  
      n_{k,z}(\rp,z) \vect e_{z} \;, \\
 \label{eq:wdvec}
    n_{k,||} (\rp,z) &=& -\ft^{-1}\bt 1^{-1} \left[ \tilde \rho(\qp,q_z)\;
    \tilde w^k_{||}(\qp,q_z) \right] \;, \\ 
 \nonumber
    n_{k,z} (\rp,z) &=& \ft^{-1}\bt 0^{-1} \left[ \tilde \rho(\qp,q_z)\;
    (-i)\tilde w^k_{z}(\qp,q_z) \right] \;. 
\eea
We demonstrate this result for the example of the weighted density $\vect n_2$.
Upon choosing $\vect r=(\rp,0,z)$ we find
\bea
  \vect n_2(\vect r) &=&  \int \frac{d\vect q}{(2\pi)^3} {\exp(-i\vect q \cdot \vect r)}
   \;\tilde \rho(\qp,z)\,(-i\vect q) \tilde w^3(q) \;, \\
   \nonumber
    &=& \int_{-\infty}^{\infty} \frac{dq_z}{(2\pi)^3} \,\exp(-iq_z z)
       \int_0^\infty \qp d\qp \,\tilde \rho(\qp,z) 
       \int_0^{2\pi} d\phi_q \,\exp(-i\qp\rp \cos\phi_q)
       \left(\begin{matrix} -i \tilde w^2_{||}\cos\phi_q \\  
                      -i \tilde w^2_{||}\sin\phi_q \\
                      -i \tilde w^2_{z} \end{matrix} \right)  \\
    &=& \int_{-\infty}^{\infty} \frac{dq_z}{2\pi} \,\exp(-iq_z z)
        \int_0^\infty \frac{\qp d\qp}{2\pi}\, \tilde \rho(\qp,z)
        \left(\begin{matrix} -J_1(\qp\rp)  \tilde w^2_{||} \\
                      0 \\
                      -i J_0(\qp\rp)\tilde w^2_{z} \end{matrix} \right) \;,
\eea
which is equivalent to Eq.~(\ref{eq:wdvec}) for $k=2$. Here we made use of
the integrals 
\begin{eqnarray*}
\int_0^{2\pi} d\phi \exp(-ix\cos\phi) &=& 2\pi J_0(x) \;, \\
\int_0^{2\pi} d\phi \exp(-ix\cos\phi) \sin\phi &=& 0 \;, \\
\int_0^{2\pi} d\phi \exp(-ix\cos\phi) \cos\phi &=& 2\pi i J_0'(x) = -2\pi i  J_1(x) \;.   
\end{eqnarray*}

Next we consider the evaluation of the convolution type integrals appearing in
$\mu[\rho]$ (see Eq.~(\ref{eq:mufunc})):
\bea
 \mu[\rho] = \sum_\alpha \int d\vect r' p_\alpha(\vect r') w^\alpha(\vect r'-\vect r) \;,
\eea
where $p_\alpha=\partial\Phi/\partial n_\alpha$ and $\alpha$ runs over scalar and
vector indices.  For scalar indices, $w^k(\vect r) = w^k(-\vect r)$:
we recover the standard convolution integral and thus:
\bea
  \int d\vect r' p_k(\vect r') w^k(\vect r'-\vect r) =
   \ft^{-1}\bt 0^{-1}\left[ \tilde p_k(\qp,q_z)\;
    \tilde w^k\left(\sqrt{\qp^2+q_z^2}\right) \right]  \qquad
   (k=0\dots 3)\;. \quad
\eea
In the case of vector indices, we observe that the free energy density $\Phi$
only depends on vector densities through $\vect n_1 \cdot \vect n_2$ and
$\vect n_2 \cdot \vect n_2$. Thus we find ($k=1,2$):
\bea
  \vect p_k(\vect r) &=& p_{k,||}(\rp,z) \vect e_{||} + 
                         p_{k,z} (\rp,z)\vect e_z  \quad \to \\
  \tilde {\vect p}_k (\vect q) &=& \tilde  p_{k,||}(\qp,q_z) \vect e'_{||} + 
                                    \tilde p_{k,z}(\qp,q_z) \vect e_z  
\eea 
with
\bea
 \tilde  p_{k,||}(\qp,q_z) = i\, \ft \bt 1 p_{k,||}(\rp,z)\;, \qquad 
 \tilde  p_{k,z}(\qp,q_z) =\ft \bt 0  p_{k,z}(\rp,z)\;.
\eea
Therefore the vector part summands of $\mu[\rho]$ are evaluated by
\bea
  \int d\vect r' p_k(\vect r') \vect w^k(\vect r'-\vect r) =
   \ft^{-1} \bt 0^{-1} \left[  -\tilde  p_{k,||} \tilde w^k_{||} + 
                               i \tilde p_{k,z} \tilde w^k_z  \right]
\eea 

\subsection{Fast Hankel transforms}

\label{sec:fht}

Hankel transforms can be calculated by employing fast Fourier transforms
on a logarithmic grid. Consider the Hankel transform
\bea
  \bt \mu f(\rp)& = & 2\pi \int_0^\infty \rp d\rp\, J_\mu (\qp \rp)\,f(\rp) \;.
\eea
We define new variables $x=\ln(\rp/r_0)$ and $y=\ln(\qp/q_0)$ with
arbitrary constants $r_0$ and $q_0$. In terms of these variables
\bea
  \bt \mu f(\rp)& = &  2\pi r_0^2 
    \int_{-\infty}^{\infty} dx\, \hat J_\mu(x+y)\, \hat f(x)\;,  
\eea 
where $\hat J_\mu(x) = J_\mu(q_0r_0 e^{x})$ and $\hat f(x)=e^{2x} f(e^x)$.
Thus, the Hankel transform takes the appearance of a cross--correlation integral
and can be solved via Fourier transforms:
\bea
  \bt \mu f(\rp)& = &  2\pi r_0^2\; \ft^{-1} \left[ \ft \hat J_\mu(x)\, 
   \ft^{\!\!*} \hat f(x) \right ]\;. 
\eea
The Fourier transform of $\hat J_\mu$ can be done analytically while the remaining
ones are computed numerically using Fast Fourier techniques.

In the actual implementation we followed Ref.~\cite{Ham00} which details
the proof of orthonormality for {\em discrete} functions, defined on a 
{\em finite} interval
in logarithmic space (and continued periodically). The following remarks apply:
\begin{itemize}
 \item The fast Hankel transform cannot be applied directly to the density
  profile, $f(\rp) \equiv \rho(\rp,z)$, since it does not go to zero for
  $\rp \to \infty$. It is therefore advantageous to split the external potential
\bea
  V^{\rm ext}(\vect r) = V(z) + V_s(\rp,z)
\eea
 into a (possibly) $z$--dependent background part $V(z)$ and the remainder. The 
 corresponding background profile $\rho_V$ fulfills
\bea
 \label{eq:rhov}
  \ln \frac{\rho_V(z)}{\rho_s} = -\beta\mu [\rho_V(z)] + \beta\mex - \beta V(z) \;. 
\eea
 The full equilibrium profile can then be written as 
 $\req(\rp,z) = \rho_V(z)(h(\rp,z)+1)$ with $h(\rp,z)\to 0$ for
 $\rp \to \infty$. The determining equation for $h$ reads
\bea
 \label{eq:h}
 \ln (h(\rp,z)+1) = - \beta\mu [\rho_V(z)h(\rp,z)+\rho_V(z)] + \beta\mu [\rho_V(z)] - \beta V_s(z)\;.
\eea
 Thus one needs to perform Hankel transforms only on functions which properly go
 to zero for $\rp \to \infty$. In the wall--sphere case, $V(z)$ is naturally given by
 the wall potential, $V_s$ becomes the solute--solvent pair potential $u_{bs}$
 and $h \equiv h_{bs}$ is the solute--solvent pair correlation function. 
 In that form, Eq.~(\ref{eq:h}) resembles the general closure for integral equations
 (Eq.~(\ref{eq:cl})). 
 In the sphere--sphere case, $V(z)=0$ with $\rho_V=\rho_s$.
 \item The assumed periodicity of $\hat f(x)$ leads to restrictions on the product
  $q_0 r_0$ (low--ringing condition in Ref.~\cite{Ham00}). This condition cannot be
  fulfilled on one grid for both $\bt 0(\cdot)$ and $\bt 1(\cdot)$. However, this
  does not lead to any noticeable instabilities in the repeated application 
  of the fast Hankel transform.
 \item In order to avoid aliasing and the amplification of numerical ``noise" in the
  low--$x$ and high--$x$ tails of $\hat f(x)$ in the repeated application of the
  fast Hankel transform we worked with cutoffs 
  $r_{\rm min}/r_0=q_{\rm min}/q_0=0.01$  and  
  $r_{\rm max}/r_0=q_{\rm max}/q_0= O(100)$. The fast Hankel transform itself
  was calculated on an extended grid $x \in [-N_{||}\Delta x/2,N_{||}\Delta x/2]$ with
  either $N_{||}=2048, \Delta x=0.01$ or $N_{||}=4096, \Delta x=0.005$. Outside
  the ``physical" domain defined by 
  $x_{\rm phys} \in [\ln (r_{\rm min}/r_0), \ln (r_{\rm max}/r_0)]$  
  the function $\hat f(x)$ was put to zero (a similar prescription applies
  to $y_{\rm phys}$).
\end{itemize}
 
\subsection{Speedup of iterations}

\label{sec:diis}

The density profile $\rho(\rp,z)$, the weighted densities $\vect n(\rp,z)$ and
the derivatives $\vect p(\rp,z)=\partial \Phi/\partial \vect n(\rp,z)$ have been
discretized on a two--dimensional grid spanning the plane $(\rp,z)$. Spacing in
$z$--direction was equidistant with grid width $\Delta z=0.002\dots 0.005\sigma$
with up to $N_z=15,000$ points. Spacing in $\rp$--direction was logarithmic
(see above) with $N_{||}\approx 1,000\dots 2,000$ points. Memory requirement went
up to 16 GB for the largest grids. Computations have been performed on
nodes with 16 GB RAM and two Intel QuadCore processors, with OpenMP 
parallelization of the arrays of Fourier and Hankel transforms. One evaluation
of $\mu[\rho]$ (one iteration) took up to two minutes.  

The equilibrium density profile fulfilling Eq.~(\ref{eq:rhoeq}) or
Eqs.~(\ref{eq:rhov}) and (\ref{eq:h}) can be determined by Picard iterations
where, as a minimum requirement for convergence, 
careful mixing of the current and previous iteration is necessary. However, for packing fractions $\eta >0.3$
and larger colloid--solvent size ratios $\alpha>5$ one easily needs several hundreds 
to thousands of iterations until convergence. In view of the iteration times
up to two minutes, this is inacceptable. Therefore we employed the modified
DIIS (direct inversion in iterative subspace) scheme developed in 
Ref.~\cite{Kov99} which essentially constructs the next iterative step out of
a certain number of previous steps. In the DIIS scheme, the mixing coefficients of the
previous steps, determining the solution of the next step, 
are obtained by a minimization  
condition on the residual. The modification to DIIS consists in the
admixture of the extrapolated DIIS residual to the solution of the next iterative
step, in order to enlarge the dimensionality of the iterative subspace and thus
to reach the true solution much more quickly. Our practical experience with modified
DIIS is very similar to the observations in Ref.~\cite{Kov99}, this includes
the necessity to combine DIIS with Picard steps carefully, in case the DIIS steps
show divergent behaviour. In summary, modified DIIS is a robust method for our 
problem, and the total number of iterations is reduced to approximately 100 and less
for most parameter choices. The only noticeable exception occured for the case of
the nearly singular annular wedge, i.e. when the distance $h$ between wall and
colloidal sphere (or the two colloidal spheres) is $\approx 1\,\sigma$ and the
solvent packing fraction $\eta_s>0.35$. Here, several hundred of intermediate Picard
steps between two DIIS steps where occasionally necessary to prepare the ground for 
the next DIIS step.

\section{Integral equations for the solute--wall case: Numerical solution }
\label{app:ie}

In cylindrical coordinates the inhomogeneous OZ
equations (\ref{eq:ozss})--(\ref{eq:ozbs}) read \bea
 \label{eq:ozapp1}
  h_{ss}(z, z_0, s) - c_{ss}(z, z_0, s) & = &
 \int d\vect s^{\prime} dz^{\prime} \rho_{V}(z^{\prime}) h_{ss}(z, z^\prime, |\vect s -\vect s^{\prime}|) c_{ss}(z^\prime, z_0, s^{\prime}) \;, \\
 \label{eq:ozapp2}
   h_{bs}(z, z_0, s) - c_{bs}(z, z_0, s) & = &
 \int d\vect s^{\prime} dz^{\prime} \rho_{V}(z^{\prime}) h_{bs}(z, z^\prime, |\vect s -\vect s^{\prime}|) c_{ss}(z^\prime, z_0, s^{\prime}) \;,
\eea where the total and direct correlation functions $h_{ij}(z,
z_0, s)$ and $c_{ij}(z, z_0, s)$ of two particles depend on the
distances $z$ and $z_0$ from the wall and on the projection of their
position vectors on the direction along the wall, $\vect s=\vect r_{||}-\vect r_{||,0}$.
The integral over $\vect s^{\prime}$ is
a 2D convolution, which is most efficiently done by means of the Fast Hankel
transform of the zeroth-order (see Appendix \ref{sec:fht}). In Fourier space, the OZ equations
become 
\begin{equation}
 \tilde h_{ij}(z, z_0,\qp) - \tilde c_{ij}(z, z_0,\qp) = \sum_{k=b,s}
  \int dz^\prime \rho_{V,k}(z^\prime)\tilde h_{ik}(z, z^\prime,\qp) \tilde c_{kj}(z^\prime, z_0,\qp)\;,
\end{equation}
where $\tilde h[c](z,z_0,\qp) = {\rm HT}_0\; h[c](z,z_0,\rp)$.
The remaining $z$-integral is evaluated with a simple
trapezoidal rule on a uniformly discretized grid of $N_z$ points,
yielding the following matrix equation:
\begin{equation}
\label{eq:ozapp3} \vect H_{ij} (\qp)-\vect  C_{ij} (\qp)=
\sum_{k=b,s}\vect H_{ik} (\qp) \mathit {\vect R} \vect C_{kj} (\qp)
\;,
\end{equation}
where $\vect H$ and $\vect C$ are $N_z\times N_z$ matrices generated
by the corresponding correlation functions at each $\qp$ point and
$\mathit {\vect R}$ is a diagonal matrix corresponding to
$\rho_V(z)\delta(z-z^\prime)$.

The Lovett--Mou--Buff--Wertheim (LMBW) equation  (\ref{eq:lmbw}) for
the background density profile in the presence of a hard wall takes
the following form:
\begin{equation}
\frac{\partial \rho_V(z)}{\partial z}= \rho_V(z)\left ( \rho_V(0)
\int d\vect\rp c_{ss}(z,0,\rp) +\int d z^\prime d\vect\rp
c_{ss}(z,z^\prime,\rp)\frac{\partial \rho_V(z^\prime)}{\partial
z^\prime}\right )\;.\label{eq:mlbw}
\end{equation}
Here the 2D integration over $\rp$ can be considered as the DC
component $\qp=0$ of the Hankel transform of the direct correlation
function $\tilde c_{ss}(z,z^\prime,\qp)$ and the remaining integral
in the second term is taken again with the trapezoidal rule. Note,
that the contact density at the plannar wall is known exactly from
the wall theorem $\rho_V(0)=\beta p$, where $p$
is the bulk pressure \cite{Henderson80}. Thus, a substitution of the contact density by
the expression for the bulk pressure provides the modified LMBW
equation for the density profile \cite{Quintana89}. Following the
ansatz of Plischke and Henderson \cite{Plischke90}, we used the
Carnahan-Starling fit for pressure, which is the first equation in
(\ref{eq:coefwbII}). This in general should improve the accuracy of
the density profiles near a plannar wall, if an approximation to the
solvent direct correlation function $c_{ss}(z,z^\prime,\rp)$ is
applied.

The closure equation (\ref{eq:cl}) is the central one in the
elaboration of a successful theory, since the OZ and LMBW equations
described above are formally exact and need to be complemented with a
third relation between between correlation total
and direct functions. Unfortunately, in its exact form this third relation, expressed
via additional bridge function $b_{ij}(\vect r, \vect r_0)$ is
defined only as an infinite sum of highly connected bridge diagrams
and so cannot be fully utilized.  In practice, one needs to resort
to different approximations, which are suited for particular
applications depending on the system state. 
We considered the Percus--Yevick (PY), Rogers--Young (RY) and modified
Verlet (MV) closures defined in Eqs.~(\ref{eq:py})--(\ref{eq:vm}).
For RY we use a
scaled one--parameter form for $\xi_{ij}$, $\xi_{ij} = \xi /(R_i+R_j)$
with $\xi=0.160$ to fulfill the single--component thermodynamic
consistency requirement. 
For $\xi=0$ one recovers the PY closure
(Eq.(\ref{eq:py})), while for $\xi\rightarrow\infty$ one obtains the HNC
closure. 
Likewise the HNC closure is recovered if $\alpha_{ij}\to\infty$
in the MV closure (Eq.~(\ref{eq:vm})).
MV is consistent with PY 
up to the fourth virial coefficient, if the $\alpha_{ij}$ are
density independent. Here we follow the suggestion of Henderson
\emph{et. al.} \cite{Henderson96} and use their definition for the
state--dependent parameters $\alpha_{ij}$.
In the infinite dilution limit considered in the present
paper these parameters reduce to the one--component form
\[
\alpha=\frac{17}{120\eta_s}\exp(-2^{\eta_s})+0.8-0.45\eta_s\;.
\]
The contact values from the MV closure proved to be in good
agreement with the values calculated from Carnahan--Starling equation
of state, which constitutes a considerable improvement over the PY and HNC
approximations.

In the application of these closures to the inhomogeneous system
the following precautions must be observed
to obtain robust results.
\begin{itemize}
 \item Technically we use the same iteration scheme as in Sec.~\ref{sec:diis} 
  for the numerical solution of Eqs.
 (\ref{eq:ozapp3})-(\ref{eq:mlbw}) and (\ref{eq:cl}) (though due to the memory restrictions 
 the $z$--grid spacing was only $\Delta z=0.05\sigma$ with up to $N_z=400$ points).
 The convergence drastically depends on the initial conditions,
 i.e. the zeroth iteration. While solving the PY closure it was enough
to initialize both solvent--solvent and solute--solvent correlation
functions with zeros, the RY and MV solute--solvent correlations have
to be initialized with a good guess to the final solution (which was
actually the corresponding PY result). The LMBW equation imposes
even stronger requirement on the initial profile, which needs to be
either close to the true profile or evolve very slowly from the flat
profile together with the corresponding correlation functions.
\item It turned out that the order of iterations is crucial for the
overall convergency: for a given approximation to the density
profile one needs to solve the closure and OZ equations prior to the
next iteration on the LMBW equation. It is therefore very costly to
get a fully self--consistent solution of all three equations and
normally the convergence goal for the background density $\rho_V(z)$ is much
lower than for the correlation functions.
\item To avoid the necessity of requiring the bulk equilibrium
behavior in correlation functions for distances far enough from the
hard wall, we introduced the second wall at $L\sim 20\sigma$ apart
from the first one. The wide slit geometry allowed us to keep the
$z$--resolution at the maximal possible, yet feasible level. 
Finer resolution in $z$ in general improves numerical stability 
of the iteration procedure.
Alternatively, one can use a very recent
method of the expansion into the orthogonal set of functions
proposed by Lado in Ref.~\cite{Lado09}.
 \end{itemize}

\end{appendix}

\end{document}